\documentclass[acmsmall,screen]{acmart}

\newif\ifarxiv
\arxivtrue

\usepackage{caption,subcaption}
\usepackage{wrapfig}
\usepackage{graphicx,fancyvrb,booktabs}
\usepackage{xspace}

\usepackage[T1]{fontenc}
\usepackage[utf8]{inputenc}
\usepackage{microtype}

\usepackage[frozencache=true,cachedir=minted-cache]{minted}

\newcommand{\iver}[1]{\mintinline{systemverilog}{#1}}
\newcommand{\iverhack}[1]{\mintinline{verilog}{#1}}

\newcommand{\gramsep}{\ensuremath{\ |\ }}

\newcommand{\orderall}{NB\_ORDER\_ALL\xspace}
\newcommand{\ordersame}{NB\_ORDER\_SAME\xspace}

\newcommand{\nomixblocking}{NB\_MIX\_NO\_BLOCKING\xspace}
\newcommand{\nomix}{NB\_MIX\_NO\xspace}

\setcopyright{cc}
\setcctype{by}
\acmDOI{10.1145/3720484}
\acmYear{2025}
\acmJournal{PACMPL}
\acmVolume{9}
\acmNumber{OOPSLA1}
\acmArticle{126}
\acmMonth{4}

\ifarxiv
\else
\received{2024-10-16}
\received[accepted]{2025-02-18}
\fi

\begin{document}

\title{The Simulation Semantics of Synthesisable Verilog}

\author{Andreas Lööw}
\orcid{0000-0002-9564-4663}
\affiliation{%
  \institution{Imperial College London}
  \city{London}
  \country{United Kingdom}
}
\email{a.loow@imperial.ac.uk}

\begin{abstract}

Despite numerous previous formalisation projects targeting Verilog, the semantics of Verilog defined by the Verilog standard -- Verilog's simulation semantics -- has thus far eluded definitive mathematical formalisation. Previous projects on formalising the semantics have made good progress but no previous project provides a formalisation that can be used to execute or formally reason about real-world hardware designs. In this paper, we show that the reason for this is that the Verilog standard is inconsistent both with Verilog practice and itself. We pinpoint a series of problems in the Verilog standard that we have identified in how the standard defines the semantics of the subset of Verilog used to describe hardware designs, that is, the synthesisable subset of Verilog. We show how the most complete Verilog formalisation to date inherits these problems and how, after we repair these problems in an executable implementation of the formalisation, the repaired implementation can be used to execute real-world hardware designs. The existing formalisation together with the repairs hence constitute the first formalisation of Verilog's simulation semantics compatible with real-world hardware designs. Additionally, to make the results of this paper accessible to a wider (nonmathematical) audience, we provide a visual formalisation of Verilog's simulation semantics.

\end{abstract}





\maketitle


\section{Introduction}


To formally reason about a programming language or a hardware-description language (HDL), a mathematical formalisation of the language is required (e.g., an operational semantics or a denotational semantics). Examples of applications of formal reasoning include the \emph{verification of programs/hardware designs} implemented in the language in question and the \emph{verification of tools} for the language, such as compilers/synthesis tools and analysis tools (e.g.,~model~checkers).

Unfortunately, the most popular HDL~\cite{Flake2020}, Verilog, lacks a definitive mathematical formalisation. The semantics of Verilog, known as its \emph{simulation semantics} (formally, its \emph{scheduling semantics}), is defined by the (System)Verilog standard, IEEE 1800-2023~\cite{SystemVerilog-2023}. As an HDL, the semantics of Verilog is distinctly different from the semantics of traditional programming languages such as C++ or Haskell. In short, the semantics is \emph{reactive} and \emph{event driven}: it is driven by \emph{concurrent processes that create and react to events}, such as a clock tick or a change in a circuit input.


Although no definitive mathematical formalisation of Verilog is available today, previous work on formalising Verilog have made good progress~\cite{Meredith10,Chen23,Gordon95,Schneider98a,Schneider98b,Pace98,Jifeng00a,Jifeng00b,Bowen00,Zhu01a,Zhu01b,Huibiao06,Stewart02}. The most comprehensive and detailed formalisation, that is, the state-of-the-art, is the formalisation by Chen~et~al.~\cite{Chen23}. Before Chen~et~al.'s work was published the state-of-the-art was the formalisation by Meredith~et~al.~\cite{Meredith10}. Despite the progress made, these previous formalisations are not definitive since they cannot be used to execute or formally reason about real-world Verilog hardware designs: in particular, the major problems remaining to be resolved are the concurrency problems at the core of Verilog's semantics that both Chen et al. and Meredith et al. run into and leave open.



In this paper, we address the concurrency problems left by previous work on formalising Verilog and, as a result, can for the first time present a formalisation of Verilog that is compatible with real-world hardware designs. We succeed in doing so by identifying problems in the Verilog standard that are the root causes of the problems left by previous work. Additionally, we also highlight problems in the Verilog standard that have gone unnoticed in previous work. Previous work have tried to formalise the standard as-is but the problems we have found in the standard suggest that this is to attack the \emph{wrong problem}: a formalisation of a broken standard is of little use. The \emph{right problem} to attack is to clearly describe the problems in the standard such that they can be resolved -- this is the problem we attack in this paper. 

\paragraph{Contribution 1: Verilog standard problems (Sec.~\ref{sec:informal-simulation-semantics}, \ref{sec:simulation-semantics}, and \ref{sec:problems})} The first contribution of this paper is that we identify a series of problems in the Verilog standard. We have found these problems in a multitude of ways. Some problems we have found by (unstructured) close reading of the Verilog standard and previous work on formalising Verilog's simulation semantics. Some of the problems we have found by developing, what we call, a ``visual formalisation'' of Verilog's simulation semantics (see the third contribution below).

The problems we identify are of varying character -- the most severe problems are problems where the Verilog standard is inconsistent with Verilog practice. In more detail, these are problems arising from inconsistencies between the following two descriptions of Verilog's simulation semantics:
\begin{itemize}
\item Verilog's simulation semantics as \emph{described by the Verilog standard}; which we call, simply, Verilog's \emph{simulation semantics} (since it is \emph{the} simulation semantics).
\item Verilog's simulation semantics as \emph{assumed in practice}; which we call, Verilog's \emph{informal simulation semantics}. This is the semantics used by, e.g., practitioners when designing hardware using Verilog and in university course material for digital design courses. 
\end{itemize}
In the main text of the paper, we discuss both descriptions and we show, for example, that Verilog's \emph{simulation semantics} allows for more interleavings between concurrent processes than Verilog's \emph{informal simulation semantics}. Verilog code written assuming Verilog's informal simulation semantics, that is, Verilog code written by today's Verilog conventions, is therefore not compatible with the semantics specified by the standard; the semantics specified by the standard, if followed to the letter, renders almost all but the most trivial Verilog code~completely~broken.


Another type of problem we highlight is where the standard is inconsistent with itself. For example, the semantics of a core concurrency construct of Verilog, called nonblocking assignments, used for race-free communication between concurrent processes, is specified in contradictory ways in different parts of the standard. The semantics of nonblocking assignments is described by both pseudocode and prose text in the standard and these two descriptions are inconsistent.






\paragraph{Contribution 2: mathematical formalisation of Verilog (Sec.~\ref{sec:formalisations})} The second contribution of this paper is that we show that the problems we have identified in the Verilog standard (the first contribution of the paper) are sufficient to enable us to ``repair'' the current state-of-the-art mathematical formalisation of Verilog, i.e., the formalisation by Chen~et~al.~\cite{Chen23}, in the sense that after the repairs the formalisation is compatible with Verilog practice. Specifically, as we detail below, we show this \emph{indirectly} by repairing Chen~et~al.'s executable implementation of their formalisation.

To be able to evaluate their formalisation on Verilog code, Chen et al. have implemented an executable version of their formalisation in Java. In their paper, Chen et al. run this Java implementation on a selection of test cases from the test suite of the open-source Verilog simulator Icarus~\cite{Icarus} and real-world test cases (including tests for parts of a CPU implemented in Verilog). Since Chen et al.'s formalisation closely follows the Verilog standard and (as we show) therefore inherits the problem of the standard, many of these tests fail to execute correctly.

We repair the problems we have identified in the Verilog standard in the Java implementation of Chen et al.'s formalisation and show that this repaired implementation can successfully execute almost all test cases Chen et al. used to evaluate their formalisation. We make the repaired implementation available in the artefact of this paper.

\paragraph{Contribution 3: visual formalisation of Verilog (Sec.~\ref{sec:formalisations})} The third contribution of this paper is that we have developed a new web-browser-based visual Verilog simulation tool VV (short for ``Verilog visualiser''). VV is the first tool to visualise the structure and maintenance of the Verilog event queue (the heart of Verilog's simulation semantics), and we therefore think of VV as the first ``visual formalisation'' of Verilog. As we explain in the body of this paper, Verilog's simulation semantics is centred around this event queue which is used to keep track of and coordinate different events. By visualising this queue, VV therethrough shows how the constructs of Verilog are given semantics in terms of their interactions with this queue. VV is interactive and driven by the user of the tool clicking the next event to execute, which allows its users to visually explore different event schedules and other aspects of the semantics of Verilog.


We believe the visual formalisation of Verilog that VV provides has complementary value to mathematical formalisation. Whereas mathematical formalisations require specialised knowledge of formal semantics to read, and hence are not accessible to a wide range of Verilog tool developers and Verilog hardware designers, our visual formalisation is readily accessible to anyone with a web browser. We believe the problems we identify here are of interest to communities outside communities where knowledge of formal semantics can be assumed and we therefore see making Verilog's simulation semantics more widely accessible through VV as an important means to disseminate our findings of this paper. Moreover, we ourselves have found VV to be helpful in debugging the Verilog standard; indeed, VV has helped us to find some of the problems of the standard we discuss~in~this~paper.

A live demo of VV and its source code are available at \url{https://github.com/AndreasLoow/vv} and also in the artefact of this paper.

\section{Scope: Target Subset of Verilog}%
\label{sec:scope}

In this section, we define the scope of our investigation into Verilog. Verilog is a large language: the Verilog standard weighs in at 1354~pages -- just the grammar alone occupies 46~pages (see App.~A of the standard). Because of the size of the standard, we have identified a \emph{target subset of Verilog} that we believe captures the subset of Verilog most important to formalise. We give the syntax of this subset in Fig.~\ref{fig:syntax}. We have identified this subset using the following three \emph{target subset selection criteria}, which we motivate and explain below:
\begin{enumerate}
\item focus on \emph{synthesisable Verilog} constructs;
\item focus on Verilog constructs that have interesting behaviour in terms of how they interact with the \emph{Verilog event queue};
\item include \emph{Verilog modules}.
\end{enumerate}

\begin{figure}[t]
\begin{minipage}[t]{0.30\textwidth}
\[
\small
\begin{array}{lcl}
  n &\in& \mathbb{N} \\
  \textit{str} &\in& \text{strings} \\
  \textit{id} &\in& \text{identifiers} \\
  b & ::= & \texttt{0} \gramsep \texttt{1} \gramsep \texttt{x} \gramsep \texttt{z} \\
  \textit{op}_\textit{1} & ::= & \texttt{!} \gramsep \texttt{\char`\~} \gramsep \cdots \\
  \textit{op}_\textit{2} & ::= & \texttt{\&} \gramsep \texttt{\&\&} \gramsep \texttt{+} \gramsep \cdots \\
%
%
  e & ::= & \texttt{\textquotesingle{}b}b \\
    & |   & \texttt{\textquotesingle{}\{}\ e{}*\ \texttt{\}} \\
    & |   & \textit{id} \\
    & |   & \textit{op}_\textit{1}\ e \\
    & |   & e\ \textit{op}_\textit{2}\ e \\
  \textit{eee} & ::= & \texttt{edge} \\
               & \gramsep & \texttt{posedge} \\
               & \gramsep & \texttt{negedge} \\
  \textit{ee} & ::= & [\textit{eee}]\ e \gramsep \textit{ee}\ \texttt{or}\ \textit{ee} \\
  \textit{st} & ::= & \texttt{\$display} \\
              & \gramsep & \texttt{\$monitor} \\
              & \gramsep & \texttt{\$finish} \\
  \textit{ste} & ::= & e \gramsep \textit{str} \gramsep \texttt{\$time} \\
\end{array}
\]
\end{minipage}
\begin{minipage}[t]{0.69\textwidth}
\[
\small
\begin{array}{lcll}
  s & ::= & s\ \texttt{;}\ s & \text{sequential sequencing} \\
    & |   & \texttt{if}\ \texttt{(} e \texttt{)}\ s\ [\texttt{else}\ s]\ & \text{if statement} \\
    & |   & \textit{id}\ \texttt{=}\ [\texttt{\#}n]\ e & \text{blocking assignment} \\
    & |   & \textit{id}\ \texttt{<=}\ [\texttt{\#}n]\ e & \text{nonblocking assignment} \\
    & |   & \texttt{@(} \textit{ee} \texttt{)}\ [s] & \text{event control} \\
    & |   & \texttt{@(*)}\ [s] & \text{comb. event control} \\
    & |   & \texttt{\#} n\ [s] & \text{delay control} \\
    & |   & \texttt{wait(} e \texttt{)}\ [s] & \text{wait statement} \\
    & |   & \textit{st}\texttt{(}\textit{ste}*\!\texttt{)} & \text{system task} \\
  m & ::= & \multicolumn{2}{l}{n \gramsep \texttt{(}n\texttt{,}\ n\texttt{)} \gramsep \texttt{(}n\texttt{,}\ n\texttt{,}\ n\texttt{)}} \\
  \textit{nt} & ::= & \multicolumn{2}{l}{\texttt{wire} \gramsep \texttt{wand} \gramsep \texttt{wor}} \\
  \textit{pt} & ::= & \multicolumn{2}{l}{\texttt{initial} \gramsep \texttt{final} \gramsep \texttt{always} \gramsep \texttt{always\_ff}} \\
              & | & \multicolumn{2}{l}{\texttt{always\_comb} \gramsep \texttt{always\_latch}} \\
  \textit{mi} & ::= & \texttt{logic}[\texttt{[}n\texttt{:}n\texttt{]}]\ \textit{id}[\texttt{[}n\texttt{:}n\texttt{]}]\ [\texttt{=}\ e] & \text{variable declaration} \\
              & |   & \textit{nt}[\texttt{[}n\texttt{:}n\texttt{]}]\ [\texttt{\#}m]\ \textit{id}[\texttt{[}n\texttt{:}n\texttt{]}]\ [\texttt{=}\ e] & \text{net declaration} \\
              & |   & \textit{pt}\ s & \text{procedure/block} \\
              & |   & \texttt{assign}\ \textit{id}\ \texttt{=}\ [\texttt{\#}m]\ e & \text{continuous assignment} \\
              & |   & \textit{id}\ \textit{id}\texttt{(}(\textit{id}.(e))*\texttt{)} & \text{module instantiation} \\
  \textit{md} & ::= & \multicolumn{2}{l}{\texttt{input} \gramsep \texttt{output} \gramsep \texttt{inout}} \\
  \textit{mk} & ::= & \texttt{logic} \gramsep \textit{nt} \\
  \textit{mp} & ::= & \multicolumn{2}{l}{\textit{md}\ \textit{mk}[\texttt{[}n\texttt{:}n\texttt{]}]\ \textit{id}[\texttt{[}n\texttt{:}n\texttt{]}]\ [\texttt{=}\ e]} \\
  m & ::= & \multicolumn{2}{l}{\texttt{module}\ \textit{id}\texttt{(}\textit{mp}*\!\texttt{)\!;}\ \textit{mi}*\ \texttt{endmodule}}
\end{array}
\]
\end{minipage}
\caption{Target syntax of expressions $e$, statements $s$, module items $\textit{mi}$, and modules~$m$. Square brackets ($[\ldots]$) denote optional elements and times ($\ldots*$) denotes repetition. Redundant syntax is omitted to avoid clutter, e.g., ``$\textit{ee} ,\,\textit{ee}$'' instead of ``$\textit{ee}\ \texttt{or}\ \textit{ee}$'' in event expressions (\textit{ee}) or \texttt{reg} instead of \texttt{logic}~in~variable~declarations~(\textit{mi}).}%
\label{fig:syntax}
\end{figure}

\paragraph{First selection criterion: synthesisable Verilog} The primary reason Verilog is a large language is because Verilog essentially consists of two languages (with shared syntax and semantics):
\begin{enumerate}
\item \emph{Synthesisable Verilog}: the subset of Verilog used to describe the structure and behaviour of hardware designs (this is the ``hardware description'' part of Verilog), deriving its name from the fact it is the kind of Verilog code that synthesis tools accept.
\item \emph{Nonsynthesisable Verilog}: the subset of Verilog used to implement ``test benches'' for hardware designs, which provide stimuli-and-probe infrastructure.
\end{enumerate}

We focus on synthesisable Verilog because our ultimate interest is formal reasoning support for Verilog: in such reasoning, test benches are replaced by other stimuli-and-probe infrastructure. E.g., in model checking the stimuli-and-probe infrastructure might be an LTL or CTL formula.\footnote{Even outside the domain of formal reasoning, some approaches to hardware development replace Verilog test benches with other infrastructure. E.g., cocotb~\cite{cocotb} enables implementing test benches in Python instead of~(nonsynthesisable)~Verilog.} With that said, we do also include a small select subset of simple nonsynthesisable Verilog in~our~target~subset. 

Fortunately, synthesisable Verilog constitutes a relatively small subset of Verilog. This is so even though we interpret ``synthesisable'' broadly in this work to protect against the fact that what kind of Verilog code is synthesisable is not officially codified anywhere. The Verilog standard does not comment on, much less codify, what kind of Verilog code is synthesisable. The previous Verilog synthesis standard~\cite{Verilog-synthesis-2005} is ``withdrawn'' without any new standard to replace it. Nevertheless, what kind of Verilog is synthesisable is relatively well-understood folklore.\footnote{Strictly speaking, what part of Verilog is synthesisable or not depends on one's synthesis tool and target technology. E.g., some target technologies support specifying initial values of registers whereas others do not. Because we interpret ``synthesisable'' broadly/generously, this does not cause any problems for us.}


\paragraph{Second selection criterion: the Verilog event queue} As we discuss further in the subsequent sections, the semantics of Verilog is oriented around an event queue, and we are especially interested in exploring a representative subset of \emph{core concurrency constructs} of Verilog that have \emph{interesting semantics in terms of how they interact with this event queue}. In particular, Verilog contains many constructs for supporting programming-in-the-large~\cite{DeRemer75} (in contrast to programming-in-the-small), such as metaprogramming constructs and modules, which are used to structure large hardware developments. Programming-in-the-large constructs are important in real-world code but not for exploring the core concurrency semantics of Verilog, this is because such constructs are typically defined by elaboration and therefore do not complicate the event queue further compared to just considering programming-in-the-small constructs.

\paragraph{Third selection criterion: modules} We make one exception from our general rule (from the second selection criterion) of not considering programming-in-the-large constructs: we consider Verilog modules. Large Verilog developments typically consist of multiple modules (representing, e.g., arithmetic logic units, memories, etc.) instantiated into large module hierarchies. Although, like other programming-in-the-large constructs, modules are given semantics by elaboration, their elaboration is interesting both from the perspective of concurrency and the Verilog event queue.

\section{Background: Informal Simulation Semantics}%
\label{sec:informal-simulation-semantics}

In this section, we introduce Verilog's \emph{informal} simulation semantics, i.e., the semantics used in everyday Verilog development. By definition, there is no authoritative source specifying the informal simulation semantics. The discussion in this section is partly based on the Verilog standard and partly based on our best understanding of the Verilog folklore.

This section serves two purposes. First, for the reader not familiar with Verilog, it serves as a crash course on the Verilog background necessary to understand the rest of the paper. Second, it makes explicit a concurrency principle, which we call CP, assumed in everyday Verilog development. When discussing problems in the Verilog standard in subsequent sections, we show that CP is not respected by the standard.

Our discussion in this section covers our target subset of Verilog (Fig.~\ref{fig:syntax}). First, we discuss the semantics of ``intermodule'' Verilog (Sec~\ref{sec:informal-intermodule-simulation-semantics}), that is, the semantics of an individual module. Second, we discuss the semantics of ``intramodule'' Verilog (Sec~\ref{sec:informal-intramodule-simulation-semantics}), that is, the semantics of a set of modules.


\subsection{Informal Intermodule Simulation Semantics}\label{sec:informal-intermodule-simulation-semantics}

In this section, we discuss the informal simulation semantics of intermodule Verilog.

\paragraph{Values.} For our target subset, Verilog values consist of bits and arrays of bits. Bits can take on four different values~\cite[p.~88]{SystemVerilog-2023}. All four are included in our target subset, see $b$ (Fig.~\ref{fig:syntax}). The value \texttt{x} is used for a multitude of purposes, often representing something like ``unknown value'', ``invalid value'', ``error'', or similar.\footnote{It is difficult to succinctly characterise precisely what role X values play in Verilog. E.g., Flake et al.~\cite{Flake2020} count eight different situations in where X values can arise.} The value \texttt{z} represents a high-impedance state and is used to model tristate logic, as discussed below.


\paragraph{Data objects.} A data object in Verilog is a ``named entity that has a data value and a data type associated with it''~\cite[p.~88]{SystemVerilog-2023}. There are two main groups of data objects in Verilog: variables and nets. Our target subset includes both groups, see $\mathit{mi}$ (Fig.~\ref{fig:syntax}). Variables have the same semantics as variables in software languages, i.e., the last write to a variable determines its value. Nets have no analogue in software languages: the value of a net is determined by its set of ``drivers'', in our target subset, a set of continuous assignments, by net resolution, as we discuss below when discussing continuous assignments.

\paragraph{Processes} Verilog is a process-based concurrent language. Our target subset includes two types of processes, discussed below: procedural processes and continuous assignments. See,~again,~$\mathit{mi}$~(Fig.~\ref{fig:syntax}).


\begin{figure}[t]
\begin{minipage}[t]{0.33\textwidth}
\begin{minted}{verilog}
logic a = 0, b = 0, c, d;

initial a = 1;

always @(a, b)
 c = a + b;

always @(c)
 d = c + b;
\end{minted}
\end{minipage}
\hfill
\begin{minipage}[t]{0.32\textwidth}
\begin{minted}{verilog}
logic a, b, clk = 0;

always #1 clk = !clk;

always @(posedge clk)
 a <= inp;

always @(posedge clk)
 b <= a;
\end{minted}
\end{minipage}
\hfill
\begin{minipage}[t]{0.33\textwidth}
\begin{minted}{verilog}
always @(*)
 c = a + b;

always_comb
 c = a + b;

always_ff @(posedge clk)
 a <= inp;
\end{minted}
\end{minipage}
\caption{Three code fragments for discussing the semantics of procedural processes.}%
\label{fig:alwaysexample}
\end{figure}

\paragraph{Procedural processes and variables} We explain procedural processes through examples. In short, procedural processes are similar to processes in traditional software languages: they come with a program counter, internal state, etc. Also like in traditional software languages, procedural processes execute in nondeterministic order. Procedural processes can only write to variables, not nets (since they cannot participate in net resolution, as discussed below).

Consider the left code fragment in Fig.~\ref{fig:alwaysexample}. The first line declares the variables \texttt{a}, \texttt{b}, \texttt{c}, and \texttt{d}, where \texttt{a} and \texttt{b} are given initial values and \texttt{c} and \texttt{d} are not and are therefore initialised with the value \texttt{x}. There are in total three procedural processes, arising from the \iver{initial} and \iver{always} blocks. An \iver{initial} block executes once and then terminates. An \iver{always} block executes over and over again in an infinite loop. The construct \iver{@(a, b)} in the first \iver{always} block is an event-control construct, which blocks the process until the value of \texttt{a} or \texttt{b} changes. The event-control construct \iver{@(c)} in the second \iver{always} block, similarly, causes the process to block until the value of \texttt{c} changes. A block can have multiple event-control constructs and they can occur anywhere in the block body. 

Blocks of \iver{always} type with only one event-control at the beginning of the block follow a simple concurrency principle, which we here call CP: such blocks are scheduled for execution each time a data object mentioned in the event-control changes. For example, this principle is useful when modelling combinational logic, that is, stateless logic. Because of CP, an \iver{always} block with only one event-control at the beginning of the block that lists all data objects the block depends on, like the first \iver{always} block (but unlike the second \iver{always} block) in our example here, is a good fit to model combinational logic. The semantics of such blocks coincide with the semantics of combinational logic because such blocks will (eventually) run with the latest values of its dependencies because such blocks are scheduled for execution each time a dependency changes.

To exemplify further, returning back to the left code fragment in Fig.~\ref{fig:alwaysexample}, given the above discussion, we see that one possible execution of the code fragment is that the \iver{initial} block runs first, updating \texttt{a}, thereby causing the first \iver{always} block to run, which in turn updates \texttt{c}, which in turn causes the second \iver{always} block to run, which updates \texttt{d}.

Now, consider the middle code fragment in Fig.~\ref{fig:alwaysexample}. This code fragment models sequential logic, that is, stateful logic. The first \iver{always} block is a behavioural model of a clock. The block uses the time-control construct \texttt{\#}\iver{1} to delay the effect of the assignment for one ``time slot'', a concept we will make more precise when introducing Verilog's simulation semantics in the next section. The two other \iver{always} blocks abide by the concurrency principle CP: i.e., the two blocks run every positive edge of the clock (\iver{posedge clk}), i.e., at every clock tick. The assignments inside the two blocks are nonblocking assignments (\iver{<=}) rather than blocking assignments (\iver{=}) as used in the left code fragment. This is to accurately model the semantics of hardware registers. In the example here, note that one block writes to \texttt{a} and another block reads from \texttt{a}. Nonblocking assignments delay the effect of their assignment until all other blocks have read the previous value; again, we will explain nonblocking assignments more precisely when introducing Verilog's~simulation~semantics~in~the~next~section.

Lastly, consider the right code fragment in Fig.~\ref{fig:alwaysexample}. The first \iver{always} block illustrates a convenient shorthand for modelling combinational logic: the block is equivalent to the first \iver{always} of the left code fragment in Fig.~\ref{fig:alwaysexample}. There are also \iver{always_comb}, \iver{always_ff}, and \iver{always_latch}, which are variants of \iver{always} which allow hardware designers to declare modelling intent (stating if the block models combinational logic, sequential logic, or latched logic, respectively). The right code fragment shows some example usages of these blocks. For our purposes here, \iver{always_comb} is the same as \iver{always_comb @(*)} except that it is guaranteed to always run at least once during the first time slot. The blocks \iver{always_ff} and \iver{always_latch} have the same semantics as \iver{always}.


\begin{figure}[t]
\begin{minipage}[t]{0.13\textwidth}
\begin{minted}{verilog}
wire w;

assign w = inp + 1;
\end{minted}
\end{minipage}
\hfill
\begin{minipage}[t]{0.33\textwidth}
\begin{minted}{verilog}
wire w1, w2;

// The value of w1 will be 1
assign w1 = 1;
assign w1 = 'z;

// The value of w2 will be x because
// the continuous assignments drive
// incompatible values
assign w2 = 0;
assign w2 = 1;
\end{minted}
\end{minipage}
\hfill
\begin{minipage}[t]{0.32\textwidth}
\begin{minted}{verilog}
wand w3, w4;

// The value of w3 will be 1
assign w3 = 1;
assign w3 = 1;

// The value of w4 will be 0
assign w4 = 1;
assign w4 = 0;
\end{minted}
\end{minipage}
\caption{Three code fragments for discussing the semantics of continuous assignments.}%
\label{fig:resolution}
\end{figure}

\paragraph{Continuous assignments and nets} We now explain continuous assignments, the second type of process we consider, again through examples. Consider the left code fragment in Fig.~\ref{fig:resolution}. The code fragment declares a net of type \iver{wire} and a continuous assignment \iver{assign} for the net. A continuous assignment induces a driver process which drives the current value of the expression of the assignment into a net or a variable. In the example, the continuous assignment drives \iver{inp + 1} into the net \iver{w}. Processes induced by continuous assignments are not like software-language processes: the only function of the driver process associated with a continuous assignment is to keep its driver value up-to-date by reevaluating the assignment expression every time a data object the expression depends on is updated. In other words, continuous assignments follow the same concurrency principle CP as simple \iver{always} blocks, although being a much simpler type of process since driver processes do not need a program counter or other internal state.

There is, however, a major difference between continuous assignments and \iver{always} blocks: continuous assignments can participate in net resolution. While a variable can have at most one driver; a net, in contrast, can have multiple drivers. If a net has multiple drivers, the values from the different drivers are merged using the resolution function of the net. Our target subset contains three types of nets: \iver{wire}, \iver{wand}, and \iver{wor}. E.g., for a \iver{wire} net, all drivers with value \texttt{z} are ignored in resolution, and all other drivers must have the same value, otherwise the net resolves to \texttt{x} -- see the middle code fragment in Fig.~\ref{fig:resolution}. A \iver{wand} net will, in contrast, resolve to the conjunction of the values of its drivers -- see the right code fragment in Fig.~\ref{fig:resolution}.

\paragraph{Other constructs} We introduce the remaining important constructs, such as \iverhack{$display} and \iverhack{$monitor}, in the next section, again by example.


\subsection{Informal Intramodule Simulation Semantics}\label{sec:informal-intramodule-simulation-semantics}

\begin{figure}[t]
\begin{minipage}[t]{0.49\textwidth}
\begin{minted}{verilog}
module circuit(
 input logic clk,
 input logic inp1,
 input logic inp2,
 output logic out);

// Model of register with xor of 
// inp1 and inp2 as input
always_ff @(posedge clk)
 out <= inp1 ^ inp2;

endmodule
\end{minted} (a) Hardware module
\newline
\begin{Verbatim}[fontsize=\footnotesize]
> iverilog -g2012 circuit.sv circuit_tb.sv
> ./a.out
time = 0 --> inp1 = x, inp2 = x, out = x
time = 1 --> inp1 = 1, inp2 = 0, out = x
time = 3 --> inp1 = 1, inp2 = 1, out = 1
time = 5 --> inp1 = 1, inp2 = 1, out = 0
circuit_tb.sv:22: $finish called at 6 (1s)
\end{Verbatim}
(c) Output from running the test bench \\ (the \texttt{-g2012} flag enables SystemVerilog \\ support)
\end{minipage}
\hfill
\begin{minipage}[t]{0.49\textwidth}
\begin{minted}{verilog}
module circuit_tb;

logic clk = 0, inp1, inp2, out;

// Instantiate the design, connecting up inputs
// and outputs by name
circuit circuit(.clk(clk), .inp1(inp1),
                .inp2(inp2), .out(out));

// Behavioural model of a clock
always #1 clk = !clk;

// Install a monitor that will print values
// (when changed) at the end of each time slot
initial begin
 $monitor("time = %0d --> ", $time,
          "inp1 = %b, inp2 = %b, out = %b",
          inp1, inp2, out);
end

// Input stimuli for the design
initial begin
 @(posedge clk) inp1 <= 1; inp2 <= 0;
 @(posedge clk) inp2 <= 1;
 @(posedge clk) @(negedge clk) $finish;
end

endmodule
\end{minted} (b) Test-bench module
\end{minipage}
\caption{Example hardware design \texttt{circuit} (a), test bench \texttt{circuit\_tb} (b), and test-bench output (c).}%
\label{fig:tbexample}
\end{figure}


In this section, we discuss the informal simulation semantics of intramodule Verilog. We again proceed by example: see the two modules in Fig.~\ref{fig:tbexample}, which consist of a hardware-design module \iver{circuit}, which is synthesisable, and a test-bench module \iver{circuit_tb}, which is not nonsynthesisable. The test-bench module \iver{circuit_tb} stimulates and probes the hardware module \iver{circuit} much like a real physical test bench would do. As shown in the example, modules can be instantiated inside other modules and connected up using the inputs and outputs of the modules. 

The contents of the two modules in Fig.~\ref{fig:tbexample} consist mostly of constructs already discussed. We mention two things of note. First, the first \iver{initial} block of the test bench installs a ``monitor'', which prints its argument at the end of each simulation time slot. In our target subset, there is also a \iverhack{$display} function that prints its argument when called instead of at the end of each time slot. Second, the second \iver{initial} block of the test bench shows that event-control constructs can occur at any location in a block.

Given a hardware design and a test bench for the design, we can simulate the design in its test bench using a Verilog simulator such as Icarus~\cite{Icarus}. Doing so using Icarus gives the output shown in Fig.~\ref{fig:tbexample}. The output shown is the output of the monitor installed by the \iverhack{$monitor} invocation in the test bench.


\section{Background: Simulation Semantics}\label{sec:simulation-semantics}


We now discuss Verilog's simulation semantics as defined by the Verilog standard. We focus on the \emph{Verilog event queue} -- the heart of the standard's description of the simulation semantics. The discussion in this section is intentionally short and only meant to give a high-level idea of the simulation semantics and introduce key terms (such as time slots and regions). We emphasise \emph{how} the standard defines the simulation semantics, which is important to understand in order to understand the problems we have found in the standard we that discuss in~subsequent~sections.

First, we discuss the intramodule simulation semantics (Sec.~\ref{sec:intramodule-simulation-semantics}). For this semantics, the standard defines the event queue and the rest of the simulation semantics by two components:
\begin{enumerate}
\item pseudocode for a ``reference algorithm for simulation'' in Sec.~4.5 of the standard, and
\item prose text at varying levels of detail (literally) sprinkled throughout the standard.
\end{enumerate}
The two components depend on each other and the pseudocode is not a complete description without the prose text and vice versa.

Second, we discuss the intermodule simulation semantics (Sec.~\ref{sec:intermodule-simulation-semantics}). This semantics is defined only by prose text. The prose text defines the intermodule semantics by elaboration to the intramodule semantics.


\subsection{Intramodule Simulation Semantics}\label{sec:intramodule-simulation-semantics}

We give a brief summary of Verilog's intramodule simulation semantics by summarising the standard's description of the reference algorithm for simulation. The reference algorithm is an interpreter for Verilog, i.e., an operational semantics. The standard describes the reference algorithm at a high level and leaves the details of the algorithm to the imagination of its readers. The entry point of the algorithm is the following pseudocode function (all pseudocode in this section is from Sec.~4.5~of~the~standard):
\begin{Verbatim}[fontsize=\small]
execute_simulation {
 T = 0;
 initialize the values of all nets and variables;
 schedule all initialization events into time zero slot;

 while (some time slot is nonempty) {
  move to the first nonempty time slot and set T;
  execute_time_slot (T);
 }
}
\end{Verbatim}
The algorithm is oriented around events and maintains an event queue. The event queue is divided into ``time slots''. The variable \texttt{T} keeps track of the current time slot, or ``simulation time''. Each time slot is split into ``regions''. The following regions are relevant for our target subset of Verilog: active, inactive, NBA (``nonblocking assignment''), and observed.\footnote{The standard includes 17 regions in total. The regions not included here are used to give semantics to constructs that are out of scope here, e.g., the Verilog APIs to interoperate with, e.g., C code.} Time slots are executed by the pseudocode function \texttt{execute\_time\_slot}. After restricting the function to the regions relevant here, the function is as follows:
\begin{Verbatim}[fontsize=\small]
execute_time_slot {
 while (any region in [Active ... Observed] is nonempty) {
  execute_region (Active);

  R = first nonempty region in [Active ... Observed];
  if (R is nonempty)
   move events in R to the Active region;
 }
}
\end{Verbatim}
That is, until all regions are empty, the events of the first nonempty region are moved to the active region and executed. The pseudocode function \texttt{execute\_region} for executing regions is as follows:
\begin{Verbatim}[fontsize=\small]
execute_region {
 while (region is nonempty) {
  E = any event from region;
  remove E from the region;

  if (E is an update event) {
   update the modified object;
   schedule evaluation event for any process sensitive to the object;
  } else { /* E is an evaluation event */
   evaluate the process associated with the event and possibly schedule
   further events for execution;
  }
 }
}
\end{Verbatim}
That is, the events of a region are executed in nondeterministic order and execution is driven by processes creating events -- update events -- and reacting to events -- evaluation events. The details of scheduling new events and executing events in the queue are described only by prose text. In short, most events are scheduled in the active region, exceptions include zero-delayed events which are scheduled in the inactive region, updates from nonblocking assignments which are scheduled in the NBA region, and \iverhack{$monitor} invocations which are scheduled in the observed region.

\subsection{Intermodule Simulation Semantics}\label{sec:intermodule-simulation-semantics}

In the Verilog standard, Verilog's intramodule simulation semantics is specified through prose-text-described elaboration (primarily in Sec. 23 of the standard, called ``Modules and hierarchy''). In this paper, we are only interested in the part of elaboration relevant to concurrency. In short, the core of the concurrency-relevant part of the elaboration boils down to the following relatively simple rules: an \iver{input} port induces a continuous assignment from outside the module to inside the module; an \iver{output} port induces a continuous assignment in the opposite direction; and an \iver{inout} port induces a ``non-strength-reducing transistor'' between the outside of the module and the inside (we discuss \iver{inout} ports further in Sec.~\ref{sec:intermodule-problems}). For example, consider again Fig.~\ref{fig:tbexample}: when the module \iver{circuit} is instantiated in the module \iver{circuit_tb}, the \iver{input} port \iver{clk} induces a continuous assignment \iver{assign circuit.clk = clk}, where the left-hand side \iver{circuit.clk} refers to the variable \iver{clk} of the instantiated module \iver{circuit}.

Examples of intermodule constructs we are not interested in here (because they do not relate to concurrency) are convenience constructs to define modules, convenience constructs to instantiate modules, and meta-programming constructs (e.g., parameterised modules and the \iver{generate} construct). For example, consider again Fig.~\ref{fig:tbexample}: using module instantiation convenience constructs, the module instantiation in the module \iver{circuit_tb} can equivalently be written as \iver{circuit circuit(.clk, .inp1, .inp2, .out)} or (the even shorter) \iver{circuit circuit(.*)}.

\section{Problems in the Verilog Standard}%
\label{sec:problems}

We now discuss the problems we have found in the Verilog standard (that previous work on formalising Verilog have not already addressed). Again, we discuss intramodule Verilog (Sec.~\ref{sec:intramodule-problems}) and intermodule Verilog~(Sec.~\ref{sec:intermodule-problems})~separately. In our discussion, we consider multiple \emph{sources of truth} and highlight discrepancies between them and the standard. Our discussion is based on the latest Verilog standard, i.e., SystemVerilog-2023~\cite{SystemVerilog-2023} (there is no separate active standard for Verilog only, Verilog has been merged into SystemVerilog).

\paragraph{First source of truth: Verilog practice} Our first source of truth is ``Verilog practice'', where we use Verilog's informal simulation semantics (outlined in Sec.~\ref{sec:informal-simulation-semantics}) and Verilog simulators as proxies for ``Verilog practice''. For our discussion on Verilog simulators, we consider both open-source and closed-source simulators. The two most popular open-source simulators are Icarus~\cite{Icarus} and Verilator~\cite{Verilator}. However, for our work here, only Icarus is relevant since Verilator does not (directly) implement Verilog's simulation semantics but instead prioritises performance over standard conformance (Verilog designs are simulated by compilation to C++/SystemC and the implemented semantics has more in common with Verilog's synthesis semantics, which is out of scope for this paper, than its simulation semantics). Our discussions are based on Icarus version 12.0. As for closed-source simulators, we consider Aldec Riviera Pro 2023.04, Cadence Xcelium 23.09, Mentor Questa 2023.3, and Synopsys VCS 2023.03.\footnote{Which are the simulators that were available at \url{https://edaplayground.com} at the time of writing.} Of course, for the internals and behaviour of closed-source simulators, we can only provide informed guesses.

\paragraph{Second source of truth: Verilog-2005} Our second source of truth is, perhaps surprisingly, the latest pre-SystemVerilog Verilog standard, i.e., Verilog-2005~\cite{Verilog-2005} (which we refer to by said name). Although the Verilog-2005 standard is officially superseded by the latest SystemVerilog standard, we believe it is meaningful to compare the two because the Verilog-2005 standard is still in use. For example, some recent work on Verilog, such as the main reference point of our work, Chen et al.~\cite{Chen23} (from 2023), published after the publication of the first SystemVerilog standard still uses the Verilog-2005 standard as its basis. The problems we highlight in this section show that even if one is only interested in a ``Verilog subset'' of SystemVerilog, one should use the latest SystemVerilog standard to avoid problems in the older versions of the standard that have been fixed in the latest version of the standard. For example, (as we show) some of the test cases that fail in Chen et al.'s evaluation of their semantics fails exactly for this reason.

\paragraph{Third source of truth: formalisations of Verilog} Our third source of truth is previous work on the formal semantics of Verilog. Although taking into consideration that they both formalise the old standard Verilog-2005~\cite{Verilog-2005}, our two main reference points are Chen et al.~\cite{Chen23} and Meredith et al.~\cite{Meredith10}, which are the two most comprehensive formalisations of Verilog.

\paragraph{Discarded sources of truth} Lastly, we want to clarify what Verilog problems we are \emph{not} interested in. Verilog, as it deserves, has a bad reputation. We believe this bad reputation is a result of, what we call, ``pragmatic shortcomings'' of Verilog; that is, commonly misunderstood aspects of Verilog (pitfalls/quirks/etc.) that arise from Verilog's, at times, antipedagogical design. Such shortcomings are relatively well understood and documented in a variety of formats, ranging from online rants (informed to varied degrees) to books such as \textit{Verilog and SystemVerilog Gotchas: 101 Common Coding Errors and How to~Avoid~Them} by Sutherland and Mills~\cite{Sutherland07}. In contrast, what we are interested in are, what we call, ``semantic shortcomings'' of Verilog; that is, inconsistencies and other types of problems that can be found in the standard that render the actual semantics of the language unclear. Whereas the pragmatic shortcomings of Verilog are relatively well understood and can be resolved by a careful reading of the standard, the semantic shortcomings we highlight in this work are novel to this work. With that said, to be clear, this is not to say that pragmatic shortcomings do not constitute a problem -- it is only to say that they are not of primary interest for our work because they are not blockers for formal reasoning.




\subsection{Intramodule Problems}\label{sec:intramodule-problems}

We discuss five intramodule problems that we have found in the Verilog standard. We call these problems PREEMPT, ALWAYS\_START, VAR\_INIT, NB\_ORDER, and NB\_MIX. We summarise how our different sources of truth handle these problems in Tab.~\ref{tab:overview}.

\begin{table}[t]
\centering
\caption{A summary of the Verilog problems we have identified in this paper. In the leftmost column, ``SV standard'' refers to SystemVerilog-2023~\cite{SystemVerilog-2023} and ``V standard'' refers to Verilog-2005~\cite{Verilog-2005}. ``--'' in any column means not mentioned. In the PREEMPT column, ``NO*'' means ``NO'' with minor exceptions (discussed in the main text of the paper). In the ALWAYS\_START column, ``NO'' means no special treatment of combinational \iver{always} blocks. For the VAR\_INIT column, see the discussion on VAR\_INIT in the main text for the meaning of INITIAL and FIRST. In the NB\_ORDER and NB\_MIX columns, NONE means that no order is enforced. In the NB\_ORDER column, ``ALL/SAME''  means that, inconsistently, both SAME and ALL are specified.}
\small
\begin{tabular}{lccccc}
 \toprule
 \textbf{Source of truth} & \textbf{PREEMPT} & \textbf{ALWAYS\_START} & \textbf{VAR\_INIT} & \textbf{NB\_ORDER} & \textbf{NB\_MIX} \\
 \midrule
 Pseudocode in SV standard & -- & -- & FIRST & NONE & NONE \\
 Prose text in SV standard & YES & NO & FIRST & ALL/SAME & BLOCK \\
 Pseudocode in V standard & -- & -- & -- & NONE & NONE \\
 Prose text in V standard & YES & NO & INITIAL & ALL/SAME & BLOCK \\
 Icarus~\cite{Icarus} & NO* & CUSTOM & FIRST & ALL & ALL \\
 Closed-source simulators & NO* & CUSTOM & FIRST & ALL & ALL \\
 Chen et al.~\cite{Chen23} & YES & NO & INITIAL & ALL & NONE \\
 Meredith et al.~\cite{Meredith10} & YES & NO & INITIAL & ALL & NONE \\
 \bottomrule
\end{tabular}
\label{tab:overview}
\end{table}

The problems PREEMPT, ALWAYS\_START, and VAR\_INIT are concurrency problems and, importantly, the main problems causing test-case failures for Chen et al. in their evaluation (see Sec.~\ref{sec:chen}). The two problems NB\_ORDER and NB\_MIX are also concurrency problems, specifically relating to the semantics of nonblocking assignments. These two problems seem to not have caused problems in previous work, instead, we found the two problems by ``visually debugging'' executions using our new visual Verilog simulator VV (which we explain further in Sec.~\ref{sec:vv} after having introduced VV).



\subsubsection{PREEMPT}


The most severe problem we have identified is that the interleaving semantics of concurrent processes specified by the standard breaks the concurrency principle CP of the informal simulation semantics (introduced in Sec.~\ref{sec:informal-simulation-semantics}), that is, is incompatible with Verilog practice. The problem, as we illustrate below, arises from the fact that the standard~\cite[p.~69]{SystemVerilog-2023} allows arbitrary interleavings by specifying that Verilog processes work by preemptive multitasking (in contrast to cooperative multitasking):\footnote{Essentially the same text has been in the standard since the very first version of the standard~\cite[p.~47]{Verilog-1995}.}
\begin{quote}
[...] statements without time-control constructs in procedural blocks do not have to be executed as one event. Time control statements are the \# expression and @ expression constructs (see 9.4). At any time while evaluating a procedural statement, the simulator may suspend execution and place the partially completed event as a pending event in the event region. The effect of this is to allow the interleaving of process execution [...].
\end{quote}


\paragraph{Minimal examples} We illustrate the problem with minimal examples. Consider the modules \iver{interleave1} and \iver{interleave2} in Fig.~\ref{fig:interleave}. The second module \iver{interleave2} is a variant of the first module \iver{interleave1} using \iver{always} instead of \iver{always_comb}, which we include here to illustrate that the problem we illustrate here happens regardless of what type of block is used. Per the discussion in Sec.~\ref{sec:informal-simulation-semantics}, the \iver{always_comb}/\iver{always} blocks of the two modules model combinational logic and should respect CP.\footnote{The \iver{initial} blocks for the purpose of this example can be seen as fragments of a combinational block (e.g., \iver{always_comb}).} In other words, according to CP, \iver{c} must equal \iver{a + b} in quiescent states. Because of the interleaving semantics specified by the standard (the text quoted above), \emph{the standard does not guarantee this}. Indeed, the following interleaving of \iver{interleave2} is allowed by the standard:
\begin{enumerate}
\item The \iver{always} process executes the \iver{@(a, b)} event control and starts waiting.
\item The \iver{initial} process executes \iver{a = 0}, which wakes up the \iver{always} process. The \iver{initial} process then preempts.
\item The \iver{always} process executes \iver{c = a + b}, which sets \iver{c} to \texttt{x} because \texttt{b} has not yet been assigned anything (and is therefore \texttt{x}), and then preempts.
\item The \iver{initial} process executes \iver{b = 1} and terminates. The \iver{always} process does not register the update event to \iver{b} since the process is not in a waiting state.
\item The \iver{always} process then continues execution by directly going into waiting since it again has reached its \iver{@(a, b)} event control.
\end{enumerate}
There are now no more events to execute, but $\mathtt{c} \neq \mathtt{a + b}$. Effectively, the \iver{always} process ``missed'' the update to the variable \iver{b}. Similar problematic interleavings can be constructed for \iver{interleave1}.

\begin{figure}[t]
\center
\begin{minipage}[t]{0.25\textwidth}
\begin{minted}{verilog}
module interleave1;

logic a, b, c;

always_comb
 c = a + b;

initial begin
 a = 0;
 b = 1;
end

endmodule
\end{minted}
\end{minipage}
\hfill
\begin{minipage}[t]{0.25\textwidth}
\begin{minted}{verilog}
module interleave2;

logic a, b, c;

always @(a, b)
 c = a + b;

initial begin
 a = 0;
 b = 1;
end

endmodule
\end{minted}
\end{minipage}
\hfill
\begin{minipage}[t]{0.25\textwidth}
\begin{minted}{verilog}
module interleave3;

logic a, b;

always_comb
 b = a;

initial begin
 a = 0;
 a = 1;
end

endmodule
\end{minted}
\end{minipage}
\hfill
\begin{minipage}[t]{0.23\textwidth}
\begin{minted}[fontsize=\footnotesize]{verilog}
// [...]

always_comb begin
 out1 = in1a;
 out2 = in2a;
 case (sel)
  cond2: out2 = in2b;
  cond3: out1 = in1c;
 endcase
end

// [...]
\end{minted}
\end{minipage}
\caption{Example Verilog code to illustrate the problem with the interleaving semantics specified~by~the~standard.}\label{fig:interleave}
\end{figure}

Another problematic example is given by the module \iver{interleave3} in Fig.~\ref{fig:interleave}. Clearly, an interleaving analogous to the problematic interleaving of \iver{interleave2} can be constructed for \iver{interleave3}. Therefore, the standard breaks \iver{interleave3} as well. Double assignments like in the \iver{initial} block happen in real Verilog code, e.g., they can arise from a common coding-style used to avoid inferring latches. E.g., the code in Fig.~\ref{fig:interleave} to the very right is example 4.5c ``case with defaults listed before case statement'' from Mills~\cite{Mills12} which illustrates this coding-style. In the code, note that both \texttt{out1} and \texttt{out2} can be assigned multiple times in the block.

\paragraph{Realistic examples}

\ifarxiv
In App.~\ref{app:interleavings},
\fi
 we provide problematic examples consisting of a synthesisable module and a test bench, to show that the problem not only occurs in artificial examples written to illustrate problems.



\paragraph{Possible fixes} The above examples show that allowing unconditional preemption breaks Verilog's informal simulation semantics. Some condition must therefore be put on when preemption is allowed. One alternative that is attractive because of its simplicity is to not allow any preemption at all, i.e., require cooperative multitasking. This change is not as disruptive as it might at first appear: as we survey below, today's simulators seem to, with minor exceptions, already implement cooperative multitasking.

\begin{figure}[t]
\center
\begin{minipage}[t]{0.35\textwidth}
\begin{minted}[fontsize=\footnotesize]{verilog}
module continterleave;

logic i, o1, o2;

// "Simple" continuous assignment
assign o1 = i;

// "Nonsimple" continuous assignment
assign o2 = i + 1;

initial begin
 $display("i = %b, o1 = %b, o2 = %b",
          i, o1, o2);
 i = 1;
 $display("i = %b, o1 = %b, o2 = %b",
          i, o1, o2);
end

endmodule
\end{minted}
\end{minipage}
\hfill
\begin{minipage}[t]{0.35\textwidth}
\begin{minted}[fontsize=\footnotesize]{verilog}
module bufinterleave;

logic i, o1, o2;

// Similar to "assign o1 = i;"
buf (o1, i);

// Similar to "assign o2 = !i;"
not (o2, i);

initial begin
 $display("i = %b, o1 = %b, o2 = %b",
          i, o1, o2);
 i = 1;
 $display("i = %b, o1 = %b, o2 = %b",
          i, o1, o2);
end

endmodule
\end{minted}
\end{minipage}
\hfill
\begin{minipage}[t]{0.2\textwidth}
\begin{minted}[fontsize=\footnotesize]{verilog}
module infiniteloop;

logic a;  
  
always
 if (a == 1) $finish;

initial
 a = 1; 
  
endmodule
\end{minted}
\end{minipage}
\caption{Two modules illustrating minor interleaving exceptions in existing simulators.}\label{fig:interleavings-simulators}
\end{figure}


\paragraph{Current simulators} First, by inspecting the source code of Icarus we see that processes are not interleaved (i.e., execute until they block),\footnote{See the comments and implementation of the main entrypoint for process execution, \texttt{vthread\_run}, at \url{https://github.com/steveicarus/iverilog/blob/v12_0/vvp/vthread.h\#L66} and \url{https://github.com/steveicarus/iverilog/blob/v12_0/vvp/vthread.cc\#L843}.} with one minor exception: simple continuous assignments may be interleaved. To illustrate that Icarus can interleave simple continuous assignments, consider \texttt{continterleave} in Fig.~\ref{fig:interleavings-simulators}. When the module is run in Icarus, we get the following output:
\begin{verbatim}
i = x, o1 = x, o2 = x
i = 1, o1 = 1, o2 = x
\end{verbatim}

Second, based on small-scale systematic testing that we have carried out (the tests and the results of the tests are included in the artefact of this paper), it appears that today's closed-source simulators do not interleave processes except for minor exceptions. To exemplify, we mention two of these minor exceptions. The first minor exception is given by Synopsys VCS: the simulator appears to interleave \iver{initial} blocks and continuous assignments. E.g., when running the module \texttt{continterleave} in Fig.~\ref{fig:interleavings-simulators} in Synopsys VCS we get the following output:
\begin{verbatim}
i = x, o1 = x, o2 = x
i = 1, o1 = 1, o2 = 0
\end{verbatim}
That is, different from the above Icarus execution of the same module, both continuous assignments are interleaved. This is a minor exception in the sense that if the \iver{initial} block is replaced with, e.g., an \iver{always_comb} block with the same body, then the continuous assignments are not interleaved. The second minor exception is given by Aldec Riviera Pro: the simulator appears to interleave \iver{buf} processes, a gate-level construct similar to continuous assignments. E.g., when the module \texttt{bufinterleave} in Fig.~\ref{fig:interleavings-simulators} is run in Aldec Riviera Pro we get the following output:
\begin{verbatim}
i = x, o1 = x, o2 = x
i = 1, o1 = 1, o2 = x
\end{verbatim}
This is a minor exception in the sense that other gate-level constructs appear to not be interleaved, e.g., as we see above, \iver{not} processes. Different from the Synopsys VCS minor exception, Aldec Riviera Pro interleaves the \iver{buf} process even if the \iver{initial} block is replaced by an \iver{always_comb} block. Lastly, if the \iver{buf} gate is replaced by an analogous continuous assignment, as specified in the comments in the module, then Aldec Riviera Pro does not interleave the block and the replacement continuous assignments.

Third, the module \texttt{infiniteloop} in Fig.~\ref{fig:interleavings-simulators} provides another way to test if today's closed-source simulators preempt processes. Aldec Riviera Pro, Cadence Xcelium, and Mentor Questa all loop forever when asked to simulate the module. Most likely, the blocks are scheduled in source code order and the first block never preempts, thereby not giving the second block a chance to execute and end the simulation. The three simulators do not loop forever when the order of the blocks is reversed in the source code. Synopsys VCS, on the other hand, never loops forever. It is unclear why: the simulation ends even if the \iver{initial} block is commented out.

\paragraph{Previous work} Both Chen et al.'s and Meredith et al.'s Verilog semantics follow the standard to the letter and allow for arbitrary interleavings. For example, both semantics allow for problematic interleaving sequences of \iver{interleave3} from Fig.~\ref{fig:interleave}. 
\ifarxiv
In App.~\ref{app:executions},
\fi
 we demonstrate this for both semantics.




\begin{figure}[t]
\center
\begin{minipage}[t]{0.32\textwidth}
\begin{minted}{verilog}
module always_start;
   
logic a, b, c;

initial a = 0;

always_comb b = a;

always @(a) c = a;

initial $monitor(a, b, c);

endmodule
\end{minted}
\end{minipage}
\hfill
\begin{minipage}[t]{0.32\textwidth}
\begin{minted}{verilog}
module var_init1;

logic a = 0;

initial $display(a);

endmodule
\end{minted}
\end{minipage}
\hfill
\begin{minipage}[t]{0.32\textwidth}
\begin{minted}{verilog}
module var_init2;

logic a;
initial a = 0;

initial $display(a);

endmodule
\end{minted}
\end{minipage}
\caption{Three example modules illustrating problems relating to the start of simulation.}\label{fig:init}
\end{figure}

\subsubsection{ALWAYS\_START} Another problem that breaks CP is the problem we call ALWAYS\_START, which relates to the first-cycle semantics of combinational blocks. To exemplify, we consider the module \iver{always_start} in Fig.~\ref{fig:init}: according to the semantics specified by the standard, this module can print \texttt{00x}, i.e., \texttt{c} can still be uninitialised at the end of simulation. This obviously breaks CP and can happen because the \iver{initial} block can execute and terminate before the \iver{always} block executes the \iver{@(a)} event control. In contrast, \texttt{b} cannot be uninitialised at the end of simulation because an \iver{always_comb} block must run at least once during the first simulation cycle. Indeed, this is one of the reasons \iver{always_comb} was introduced in the language. Introducing a new block type does, however, not solve the original problem: since \iver{always} blocks cannot be used to soundly model combinational logic, the construct should have been deprecated for this usage when \iver{always_comb} was introduced.

To work around the ALWAYS\_START problem, Icarus implements a custom execution ordering that ensures that combinational \iver{always} blocks enter waiting before other blocks start executing.\footnote{See \url{https://github.com/steveicarus/iverilog/blob/v12_0/elaborate.cc\#L6064}.} The closed-source simulators we consider, except Cadence Xcelium, seem to implement a similar approach: they all print \texttt{000}. Cadence Xcelium prints \texttt{00x} and prints \texttt{000} when the \iver{initial} block is moved after the \iver{always} block. Both Chen et al.'s and Meredith et al.'s semantics allow the~output~\texttt{00x}.

\subsubsection{VAR\_INIT} Although not a problem of the most recent standard, the problem we call VAR\_INIT has caused problems in previous work on formalising Verilog (discussed below). To illustrate the problem, consider the module \iver{var_init1} in Fig.~\ref{fig:init}. According to the most recent standard, this module can only print \texttt{0}. This is because initialisation is defined to happen before any process begins (see the pseudocode in Sec.~\ref{sec:simulation-semantics}). In Tab.~\ref{tab:overview}, we call this semantics FIRST. In Verilog-2005 (see Sec.~6.2.1 of the standard), on the other hand, variable initialisation is defined by elaboration to an \iver{initial} block with an assignment: i.e., \iver{var_init1} would be elaborated to the module \iver{var_init2} in Fig.~\ref{fig:init}. In Tab.~\ref{tab:overview}, we call this semantics INITIAL. In this earlier semantics, the module \iver{var_init1} has a race: the module can print either \texttt{1} or \texttt{x} since the elaborated \iver{initial} block can execute either before or after the display block.

This change in the semantics appears to not be well-known: Chen et al. implement the (broken) Verilog-2005 semantics. Seemingly unaware of VAR\_INIT, they misattribute resulting failures in test cases to races in the test cases.

\subsubsection{NB\_ORDER}

The semantics of nonblocking assignments specified by the pseudocode of the reference algorithm for simulation (discussed in Sec.~\ref{sec:simulation-semantics}) and the semantics specified by the prose text in the standard are not consistent with each other: the reference algorithm does not offer the same order guarantees for nonblocking assignments as the prose text. Whereas the function \texttt{execute\_time\_slot} of the pseudocode suggests that executing NBA events is a simple matter of moving all NBA events from the NBA region to the active region, the prose text of the standard suggests that executing NBA events is more involved.

\begin{figure}[t]
\center
\begin{minipage}[t]{0.32\textwidth}
\begin{minted}{verilog}
module nbinterleave1;

logic a;

initial
 $monitor(a);

initial begin
 a <= 0;
 a <= 1;
end
  
endmodule
\end{minted}
\end{minipage}
\hfill
\begin{minipage}[t]{0.32\textwidth}
\begin{minted}{verilog}
module nbinterleave2;

logic a;

always @(*)
 $display(a);

initial begin
 a <= 0;
 a <= 1;
end
  
endmodule
\end{minted}
\end{minipage}
\hfill
\begin{minipage}[t]{0.32\textwidth}
\begin{minted}{verilog}
module nbinterleave3;

logic a, b;

always @(*)
 $display(a, b);

initial begin
 a <= 1;
 b <= 1;
end
  
endmodule
\end{minted}
\end{minipage}
\caption{Three example modules illustrating problems relating to nonblocking assignments.}\label{fig:nbproblem}
\end{figure}

A defining characteristic of the active region is that its events are allowed to execute in any order, hence, doing what the pseudocode suggests and simply move all NBA events to the active region does not guarantee any order between the NBA events when executed. This is inconsistent with the prose text, which provides order guarantees. Which order guarantees should be provided is, however, unclear since the prose text is inconsistent with itself and specifies two different order guarantees. On page~69 of the standard, an order guarantee we call NB\_ORDER\_ALL is given: NBA events, \emph{unconditionally}, ``shall be performed in the order the [nonblocking assignments] were executed''. On page~254 of the standard, an order guarantee we call NB\_ORDER\_SAME is given: ``[t]he order of the execution of distinct nonblocking assignments to a given variable shall be preserved'', i.e., only the order of NBA events \emph{for the same variable} is required to be preserved. To exemplify, consider the module \iver{nbinterleave1} in Fig.~\ref{fig:nbproblem}. Enforcing either NB\_ORDER\_ALL or NB\_ORDER\_SAME, the module must print \texttt{1}. Enforcing neither, the module can print either \texttt{0} or \texttt{1}. Clearly, one of NB\_ORDER\_ALL or NB\_ORDER\_SAME must be enforced: much of today's Verilog code would break if \iver{nbinterleave1} would be allowed to print \texttt{0}.

We discuss how our different sources of truth handle this problem after having introduced the next problem, NB\_MIX, since the two problems are interrelated.

\subsubsection{NB\_MIX} The prose text of the standard enforces another order guarantee of NBA events not enforced by the pseudocode. The following prose text can be found on page~253 of the standard, specifying an order guarantee we call \nomixblocking:\footnote{Another unrelated problem: note that the quoted text has not been updated for SystemVerilog, specifically, assignments in the reactive region set execute after assignments in the active region set. The same text occurs in the Verilog-2005 standard~\cite[p.~120]{Verilog-2005}. This is not of great importance here since the reactive region set is out of scope for our work.}
\begin{quote}
[...] nonblocking assignments are the last assignments executed in a time step---with one exception. Nonblocking assignment events can create blocking assignment events.~These blocking assignment events shall be processed after the scheduled~nonblocking~events.
\end{quote}
I.e., the execution of blocking and nonblocking assignments must not be mixed. As far as we can tell, the standard does not comment on other types of events mixing with NBA events. E.g., is the execution of the \iverhack{$display} statement in \iver{nbinterleave2} in Fig.~\ref{fig:nbproblem} allowed to mix with the execution of the nonblocking assignments? That is, should it be possible for the module to print twice or just once? For later discussion, we call the order guarantee that disallows all mixing \nomix.

\paragraph{Possible fixes} For NB\_ORDER, at least for the subset of Verilog we consider here, enforcing \nomix, which we believe should be enforced per the below discussion, resolves the dilemma of which of \orderall and \ordersame should be enforced: given \nomix, \orderall and \ordersame have no externally observable difference between them. To exemplify, consider \iver{nbinterleave3} in Fig.~\ref{fig:nbproblem} and assume that \nomix is enforced: now, we see that all possible executions under both \orderall and \ordersame give the same output.

For NB\_MIX, what order guarantee should be enforced is less clear: should no order guarantee, \nomixblocking, or \nomix be enforced? We believe there are advantages to enforcing \nomix: this is because enforcing \nomix makes the interleaving behaviour of nonblocking assignments consistent with the interleaving behaviour of procedural processes. We illustrate by example: consider again \iver{nbinterleave2} and \iver{nbinterleave3} from Fig.~\ref{fig:nbproblem}. With either no order guarantee or \nomixblocking, the modules \iver{nbinterleave2} and \iver{nbinterleave3} are able to invoke the \iverhack{$display} call multiple times. In contrast, if all the nonblocking assignments of \iver{nbinterleave2} and \iver{nbinterleave3} are replaced by blocking assignments, then the modules would only be able to invoke \iverhack{$display} at most once. That is, nonblocking assignments under no order guarantee or \nomixblocking have more interleavings than the corresponding blocking assignments would. Enforcing \nomix rules out such ``extra'' interleavings, making the interleaving behaviour of nonblocking assignments and procedural processes more consistent.

\paragraph{Current simulators} Icarus enforces both NB\_ORDER\_ALL and \nomix.\footnote{See \url{https://github.com/steveicarus/iverilog/blob/v12_0/vvp/schedule.cc\#L1119}.} Icarus moves all NBA events from the NBA region to the active region when the active region and inactive region are empty, exactly as specified by the pseudocode of the standard. Despite simply moving the NBA events to the active region, because Icarus executes all events in the active region in FIFO order, the simulator enforces both \orderall and \nomix.

All closed-source simulators we have tested give the same output for the three modules from Fig.~\ref{fig:nbproblem}, specifically, \texttt{1} for \iver{nbinterleave1}, \texttt{1} for \iver{nbinterleave2}, and \texttt{1, 1} for \iver{nbinterleave3}, indicating that no reordering or mixing has taken place. We have not managed to find an example of where an existing simulator reorders NBA events or mixes them with other events.

\paragraph{Previous work} Neither Meredith et al. nor Chen et al. highlight the two nonblocking-assignment inconsistencies we highlight here. Nevertheless, both Meredith et al.'s and Chen et al.'s semantics enforce \orderall. Meredith et al.'s semantics, when moving NBA events in the NBA region to the active region, maintain the order among the NBA events by grouping all NBA events into a group event containing all the NBA events and move this group event to the active region instead of each individual NBA event. In the active region, the events contained inside NBA group events are executed in order. Chen et al. take a bigger step away from the pseudocode and maintain the order of NBA events by spawning a new procedural process containing all writes of the NBA region instead of moving events between regions; we have found no support for this behaviour in the standard. None of \nomixblocking and \nomix are maintained by Meredith et al.'s or Chen et al.'s semantics. 
\ifarxiv
In App.~\ref{app:executions},
\fi
 we demonstrate using examples how Meredith et al.'s and Chen et al.'s semantics handle nonblocking assignments.





\subsection{Intermodule Problems}\label{sec:intermodule-problems}

We discuss two intermodule problems we have found in the Verilog standard. These two problems are not as severe as the intramodule problems discussed in the previous section, but should nevertheless be addressed in the standard. Both problems relate to Chen et al.'s previous work on formalising Verilog. Although Chen et al., like all other previous work on Verilog, do not capture the intermodule semantics formally, they capture the intermodule semantics in their Java implementation of their semantics. The second problem we discuss here (relating to port coercion) shows up in their evaluation of their semantics.









\subsubsection{Semantics of \iver{inout} Ports}

One part of the intermodule semantics of Verilog that seems to have caused confusion in earlier work on formalisation is the semantics of \iver{inout} ports. Chen et al. claim they have ``discovered'' what they call the ``alias semantics'' of \iver{inout} ports: they handle ``inout port connections by replacing all occurrences of aliased nets with a fresh uniformed name''. This, however, is different from how the standard specifies the semantics of \iver{inout} ports -- which Chen et al. appear to have missed. This indicates that the standard should be clarified on this point. For our target subset of Verilog, a port for a module can be either an \iver{input} port, an \iver{output} port, or an \iver{inout} (bidirectional) port. Whereas the standard specifies that \iver{input} ports and \iver{output} ports are elaborated to continuous assignments, an \iver{inout} port is instead elaborated to a bidirectional connection ``analogous to an always-enabled \iver{tran} connection between the two nets, but without any strength reduction''. (We do not cover strength reduction in this paper, see the standard for more information.)


\begin{figure}[t]
\center
\begin{minipage}[t]{0.49\textwidth}
\begin{minted}{verilog}
module coercion_in(
 input wire inp,
 output wire outp);

 assign outp = 0;
 assign inp = outp;

endmodule

module coercion_in_top;

 wire inp, outp;

 coercion_in m(.*);

 assign inp = 1;

 initial $monitor(inp, outp);

endmodule
\end{minted}
\end{minipage}
\hfill
\begin{minipage}[t]{0.49\textwidth}
\begin{minted}{verilog}
module coercion_out(
 output wire outp);

 assign outp = 0;

 initial $monitor(outp);

endmodule

module coercion_out_top;

 wire outp;

 coercion_out m(.*);

 assign outp = 1;

endmodule
\end{minted}
\end{minipage}
%
%
\caption{Intermodule simulation semantics problems.}\label{fig:intermodule}
\end{figure}

\subsubsection{Port Coercion} A second thorny issue is, what the standard calls, port coercion. The standard specifies that ``[a] port that is declared as input (output) but used as an output (input) or inout may be coerced to inout. If not coerced to inout, a warning shall be issued.'' What is meant by ``used as'' an input or output is not clear from this text. We illustrate by example: consider the left subfigure in Fig.~\ref{fig:intermodule}. In the subfigure, the \iver{input} port of \iver{coercion_in} must be coerced into an \iver{inout} port since the module writes to it. However, the \iver{output} port of the same module does \emph{not} need to be coerced even though the module is reading from it. This is because reading an output port is different from the port being ``used as'' an input port. To contrast, the right subfigure in Fig.~\ref{fig:intermodule} illustrates using an \iver{output} port as an \iver{input} port -- i.e., the write should come from outside the module. All the simulators we consider print \texttt{x} when running \iver{coercion_out_top}, indicating that the continuous assignment outside the module reaches inside the module despite the \iver{outp} port being declared an \iver{output} port. Unfortunately, Chen et al. get the distinction between ``used as'' an input port and reading a port inside a module wrong (see the discussion in Sec.~\ref{sec:chen}). The standard should therefore be clarified on this point.




\section{Verilog Formalisations}%
\label{sec:formalisations}

In this section, first, we discuss how we have repaired Chen et al.'s semantics based on the problems in the Verilog standard we have pinpointed in the previous section, second, we introduce our new visual Verilog simulation tool VV.


\subsection{Mathematical Formalisation: Repaired Chen et al. Semantics}%
\label{sec:chen}

To test their semantics, Chen et al. run their Java implementation of their semantics on a selection of test cases from the Icarus test suite and a small collection of real-world test suites, including tests for parts of a RISC-V processor. Chen et al. run into problems with test cases from both types of test suites. For the Icarus tests, Chen et al. report 116 failing tests out of 824 tests~\cite[Fig.~1]{Chen23b}. For the real-world tests, some of the tests only work under their semantics in the sense the test cases \emph{can} pass, by running the test cases multiple times until an appropriate schedule happens to be selected, rather than passing every run (which is what happens when the tests are run in using a standard Verilog simulator).

Chen et al.'s evaluation results can be significantly improved with small changes to the Java implementation. Directed by the discussion in Sec.~\ref{sec:problems}, we have modified the Java implementation (as documented in the artefact of the paper) to disable process preemption, ensure that combinational \iver{always} blocks execute at least once in the first time slot (like \iver{always_comb}), and to implement SystemVerilog semantics for variable initialisation. Below, we discuss the results of re-running the test suites with this updated Java implementation.

\paragraph{Icarus test suite} Re-running the Icarus test suite, only 23\footnote{We have removed the test \texttt{sqrt32.v} identified as broken by Chen et al. and reverted changes to \texttt{pr1913918b.v} and \texttt{pr1777103.v} by Chen et al. relating to port coersion.} tests now fail (compared to 116 before). The failing tests are summarised in Tab.~\ref{tab:test-cases} and we explain in more detail below why the remaining failing tests do not indicate any major remaining problems in the repaired semantics for executing hardware designs (i.e., synthesisable Verilog). In the below discussion, the ``(five) simulators'' we refer to are the five simulators introduced in Sec.~\ref{sec:problems}.

\begin{wraptable}{r}{0.35\textwidth}
\caption{Classification of failing tests, see further discussion in main text.}
\begin{tabular}{lc}
 \toprule
 \textbf{Type of failure} & \textbf{Num.} \\
 \midrule
 Start race & 4 \\
 Order race & 2 \\
 Other race & 3 \\
 Port coersion & 5 \\
 Net delay initialisation & 1 \\
 PCA & 8 \\
 \bottomrule
\end{tabular}
\label{tab:test-cases}
\end{wraptable}


\begin{figure}[t]
\center
\begin{minipage}[t]{0.32\textwidth}
\begin{minted}{verilog}
module almost_comb;

logic a, b;
  
initial a = 0;

always @(*) b <= #1 a;
   
initial $monitor("a = ", a,
		 ", b = ", b);

endmodule
\end{minted}
\end{minipage}
\hfill
\begin{minipage}[t]{0.32\textwidth}
\begin{minted}{verilog}
module fifo;

logic[1:0] a;

initial begin
 #5 a = 1; #5 a = 3;
end

initial #10 a = 2;

initial $monitor("a = ", a);

endmodule
\end{minted}
\end{minipage}
\hfill
\begin{minipage}[t]{0.32\textwidth}
\begin{minted}{verilog}
module net_delay;

wire w;

assign #10 w = 0;

initial $monitor("w = ", w);

endmodule
\end{minted}
\end{minipage}
\caption{Simplified failing test cases.}%
\label{fig:test-cases}
\end{figure}



\paragraph{Icarus test suite: race problems} Race problems have the potential to be interesting since our main focus is concurrency. However, none of the tests failing because of race problems are synthesisable (e.g., uses delays) and are, hence, out of immediate interest -- we nevertheless discuss them briefly below.


The five tests grouped under ``start race'' (\texttt{pr1570451b.v}, \texttt{pr1956211.v}, \texttt{pr2470181b.v}, and \texttt{pr2597278.v}) fail because of races at the start of simulation. Consider the module \texttt{almost\_comb} in Fig.~\ref{fig:test-cases}, which we have constructed to illustrate the common problem among the failing test cases. Should \texttt{b} in the last output from the monitor be allowed to be \texttt{x}? (Cadence Xcelium does allow this output, the other four simulators appear not to.) In other words, should there be a principle that ensures that the \iver{always} block in the example starts waiting before the first \iver{initial} block runs? We do not discuss further since our immediate interest here~is~synthesisable~code.


The two tests grouped under ``order race'' (\texttt{sdw\_dsbl.v} and \texttt{vardly.v}) succeed in all five simulators but fail in Chen et al.'s semantics. Consider the module \texttt{fifo} in Fig.~\ref{fig:test-cases}, again constructed to illustrate the common problem among the failing test cases. Must \texttt{a} in the last output from the monitor be \texttt{3}? (All five simulators' last output is \texttt{a = 3}.) This is only ensured if the events of the active region are executed in first-in-first-out order. This is a very strong order guarantee and indicates that the tests lack sufficient synchronisation.






The three tests grouped under ``other race'' (\texttt{br991a.v}, \texttt{pr1662508.v}, and \texttt{undefined\_shift.v}) clearly lack sufficient synchronisation. E.g., both \texttt{br991a.v} and \texttt{pr1662508.v} fail even in Icarus if the source code order of the blocks of the tests are changed.


\paragraph{Icarus test suite: port coersion problems} Five tests fail because of problems related to port coersion (grouped under ``port coersion''). Three of the tests (\texttt{bufif.v}, \texttt{pr1777103.v}, and \texttt{pr1913918b.v}) fail because \iver{output} ports are read, which Chen et al.'s semantics incorrectly disallows (recall the discussion in Sec.~\ref{sec:intermodule-problems}). The tests pass in all five simulators. All tests pass in Chen et al.'s semantics when refactored to not read from \iver{output} ports. The test \iver{br1001.v} fails because Chen et al. have chosen to not implement Sec.~23.3.3.7 of the standard. Lastly, the test \iver{gate_connect1.v} fails because array expressions are used as inputs to primitive gates. As the comment in the test file says, ``[t]he standard is quiet about [array expressions as inputs for primitive gates], but the consensus among other simulators is that the [least-significant bit] of the expression is used.'' Indeed, the test passes in all five simulators.


\paragraph{Icarus test suite: net delay initialisation problem} The single test of the group ``net delay initialisation'', \texttt{ldelay1.v}, passes in all five simulators but fails in Chen et al. semantics. We illustrate the problem using a simplified example, see \iver{net_delay} in Fig.~\ref{fig:test-cases}. As Chen et al. highlight, the standard should be more clear on what the value of the net \texttt{w} should have before time step 10, specifically, the value driven the \iver{assign} of the example. Whereas Chen et al. semantics prints ``\texttt{w = z}'' followed by ``\texttt{w = 0}'', there appears to be consensus in Verilog practice that the initial driven value should be \texttt{x}, indeed, all five simulators print ``\texttt{w = x}'' followed by ``\texttt{w = 0}''.

\paragraph{Icarus test suite: PCA problems} The tests grouped under ``PCA'' use procedural continuous assignments~\cite[Sec.~10.6]{SystemVerilog-2023}, which are outside our target scope and the tests are therefore not of immediate interest to us. See Chen et al.~\cite[Sec.~3.2.2]{Chen23b} for more discussion.

\paragraph{Real-world test suite} Re-running the real-world tests, all tests now appear to pass every run. Unfortunately, because Chen et al.'s state space exploration tool does not scale, it is only possible to run the real-world tests by running them multiple times with randomly selected schedules rather than exhaustively enumerating the entire reachable state space.

\paragraph{Conclusion} In conclusion, with only minor changes to Chen et al.'s Java implementation, the implementation is turned into an implementation compatible with Verilog practice and real-world hardware designs (as represented by the test suites included in the evaluation). Here, we have only worked with the Java implementation of Chen et al.'s semantics since that is the version of the semantics needed to run the evaluation. Applying the analogous simple fixes to Chen et al.'s mathematical formalisation (i.e., the inference rules presented in their paper) turns the formalisation into the first formalisation compatible with Verilog practice.

\subsection{Visual Formalisation: VV}%
\label{sec:vv}



We now discuss VV, the new visual Verilog simulation tool we have built. Below, we discuss the interface of VV, give some comments on the implementation of VV, and give an example of how VV has been useful to us to ``visually debug'' the Verilog standard. VV implements and visualises Verilog's simulation semantics, including the reference algorithm for simulation and the algorithm's associated event queue (discussed in Sec.~\ref{sec:simulation-semantics}). VV supports our target subset of Verilog (Fig.~\ref{fig:syntax}), except modules and arrays (which we found would not make VV's implementation of the event queue more interesting). 





\begin{figure}[t]
\centering
\includegraphics[width=\textwidth]{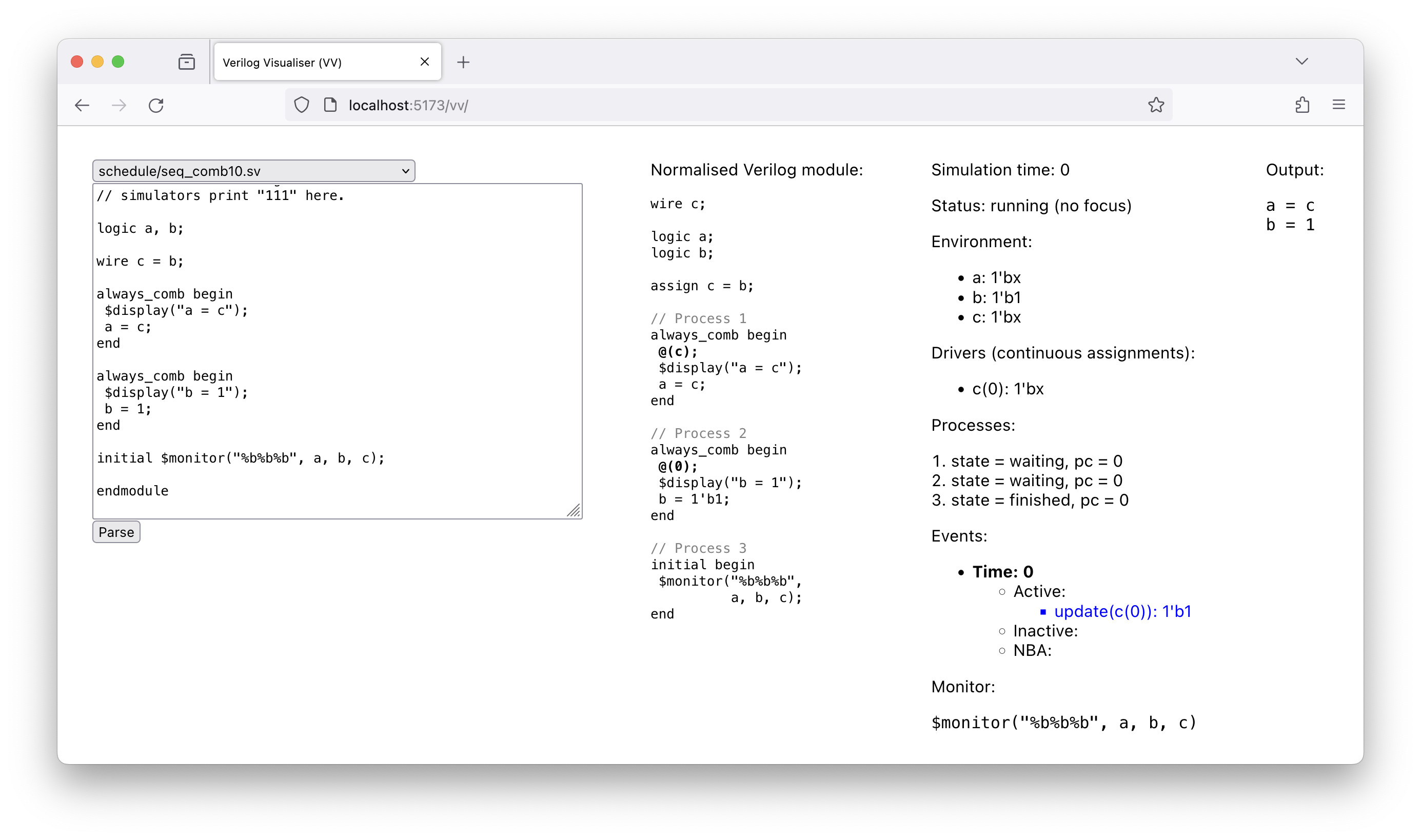}
\caption{A screenshot of VV.} 
\label{fig:screenshot}
\end{figure}

\paragraph{The interface of VV} Fig.~\ref{fig:screenshot} contains a screenshot of VV. The interface of VV is as follows, from left-to-right:
\begin{enumerate}
\item The first column of the interface contains a drop-down menu with a small collection of test modules (at the time of writing, approximately 130 modules, which we have created during the development of VV) and the source code of the currently selected test module -- the source code can also be manually edited instead of loading a predefined test module.
\item The second column contains the normalised result of parsing the module in the first column.
\item The third column contains the current state of the simulation. In more detail, from top-to-bottom: the current simulation time (and simulation status), the current state of all variables and nets, the current state of all continuous assignments (i.e., net drivers), the current state of all procedural processes, the current event queue, and the currently~installed~monitor~(if~any).
\item The fourth column contains the module output of the run so far (from, e.g., \iverhack{$display} and \iverhack{$monitor} calls).
\end{enumerate}

Simulation in VV is driven by the user clicking the next simulation step to happen. Possible next steps for the simulation are marked in blue in the third column of the interface, e.g., an event in the event queue ready to execute or the simulation-time text when the current time slot is empty and the simulation is ready to progress to the next nonempty time slot. E.g., in the screenshot in Fig.~\ref{fig:screenshot}, there is one blue-marked active event. Clicking the blue event causes VV to execute the event and update the event queue and other simulation state accordingly. After executing the clicked event, the simulator goes back into waiting for the next user decision. If needed, e.g., when there are multiple events to choose from, execution can be restarted by re-parsing the module.


\paragraph{Implementation of VV} VV can be run in any web browser without any setup needed. We have implemented VV in ReScript~\cite{rescript}, an OCaml dialect of JavaScript. We have used React~\cite{react} for the front-end of VV and Ohm~\cite{ohm} for parsing Verilog source code. The most important part of VV is its implementation of its simulation-state representation, in particular, the Verilog event queue. We have tried to capture the description of the standard, taking into consideration both previous work on formalising Verilog and the problems we have pinpointed in this paper. We describe the simulation-state representation of VV in more detail in 
\ifarxiv
App.~\ref{app:vv}.
\fi
Regarding the problems we have highlighted in this paper: VV does not allow preemption and enforces both NB\_ORDER\_ALL (using the trick by Meredith et al. described earlier in the paper) and NB\_MIX\_NO. Other implementation details of VV are not particularly interesting; the rest of VV is a simple and straightforwardly implemented event-driven interpreter, which required no particular ingenuity to implement. 


\paragraph{How VV has been useful to us} One example of how VV has been useful to us is the problem NB\_MIX. The reason we discovered this problem is that we were executing different test modules in VV (essentially playing around with the tool), and two of them happened to be the module \texttt{nbinterleave2} from Fig.~\ref{fig:nbproblem} and the analogous module with blocking assignments. Using a version of the tool that did not enforce any order between nonblocking events and other active events, the tool's visualisation of the simulation state in combination with the tool's ability to interactively execute modules made it immediately clear that the nonblocking-assignment version of the module allowed for more interleavings than the blocking-assignment version of the module, which we thought was strange and led us to investigate what order guarantees the standard specifies, which in turn ultimately led us to discover the problem we now call NB\_MIX.

\section{Related Work}\label{sec:related-work}


We discuss related work, including previous Verilog formalisations and visualisations.



\paragraph{Formalising language standards} Running into problems in language standards is not as uncommon as one might expect. In theory, a standard is the ultimate authority on the semantics of the language it standardises. In practice, less so -- as previous work exemplifies, formalising language standards is rarely as straightforward as theory suggests it should be. Memarian et al.~\cite{Memarian16,Memarian19}, in their work on formalising C, argue and give examples showing that the practice of C (what they call the ``de facto standards'' of C) and the ISO standard for C are out of sync with each other; or, in their own words: ``properties of C assumed by systems code and those implemented by compilers have diverged, both from the ISO standards and from each other, and none of these are clearly understood.'' To address this problem, they consult and balance an eclectic collection of sources of truth: the ISO standard, existing C code, experimental data from compilers, and survey and interview answers about C from systems programmers and compiler writers. These kinds of problems are not unique to C: Bodin et al.~\cite{Bodin14}, in their work on formalising JavaScript (arguably a much simpler language than C), to make sense of the semantics of JavaScript, had to consult not only the JavaScript standard but also browser implementations of JavaScript, discussion groups such as es-discuss, and the official ECMA test suite test262.

\paragraph{Verilog formalisations.} Chen et al.'s~\cite{Chen23} Verilog semantics is to date the most complete formalisation of the Verilog standard, specifically, the Verilog-2005 standard~\cite{Verilog-2005}. Chen et al. present their semantics by inference rules and also implement their semantics in Java, e.g., they have built a state-space explorer. Before Chen et al.'s semantics, the most complete formalisation was the formalisation by Meredith et al.~\cite{Meredith10}. Meredith et al.'s semantics is implemented in the K framework~\cite{Rosu10}, which allows for the generation of, among other tools, interpreters and model checkers. Gordon's~\cite{Gordon95} early work on Verilog semantics is another project of note. The project covers many important Verilog features, such as nonblocking assignments and delays. The presentation of Gordon's semantics is, however, informal (and, in places, nonstandard): the semantics is presented in prose form. Lastly, the work-in-progress paper by Lööw~\cite{Loow22b} reports on early work on the results presented in this paper.







Other previous projects on the semantics of Verilog we are aware of do not follow the standard as closely as Chen et al., Meredith et al., and Gordon. We consider those projects to be either suggestions for alternative semantics for Verilog (rather than formalisations of the standard semantics) or semantics derived from the standard semantics designed to aid formal reasoning. 


\paragraph{Verilog visualisations.} Our new visual simulator VV visualises Verilog's simulation semantics, including the event queue of Verilog. To our best knowledge, no existing Verilog simulator allows for its event queue to be inspected, whereas in VV it is the main function of the tool.\footnote{Not even the Verilog APIs (PLI/VPI) defined by the standard that allow ``foreign language functions to access the internal data structures of a SystemVerilog simulation''~\cite[Ch.~36]{SystemVerilog-2023}, do, as far as we are aware, allow for such inspection.} Analysis/debugging facilities of existing simulators are designed with the aim to help its users to find and understand bugs in Verilog designs rather than visualising Verilog's simulation semantics. Common debugging facilities in existing simulators include ``printf debugging'' (e.g., \iverhack{$display} and \iverhack{$monitor}) and waveform visualisation.

Looking beyond simulators, we are not aware of any nonsimulator tool visualising Verilog's simulation semantics. For Verilog's synthesis semantics, Materzok~\cite{Materzok19} has developed DigitalJS, ``a visual Verilog simulator for teaching''. DigitalJS uses the synthesis tool Yosys~\cite{Yosys} for synthesising its Verilog input and visualises the synthesised output of Yosys. However, the visualisation does not explain the internals of the synthesis process -- the synthesis tool is still a black box for the user (i.e., it is only its final output that is visualised). Similar visualisations, although not interactive, come bundled with e.g. Yosys itself and other synthesis tools like Xilinx Vivado~\cite{vivado}. (On the topic on visualising synthesis algorithms, although not directly related to Verilog, Nestor~\cite{Nestor08} has implemented CADAPPLETS (later ported to CADApps) for visualising a selection of synthesis algorithms.)

\section{Conclusion}%
\label{sec:conclusion}

Our aim of the project presented in this paper has changed during the course of the project. Initially, our aim was to develop mechanised metatheory for Verilog's simulation semantics such that we would be able to formally connect up Verilog tools using Verilog's simulation semantics (e.g., Verilog simulators) and Verilog tools using Verilog's synthesis semantics (e.g., Verilog synthesis tools). However, we quickly realised that the simulation semantics was too broken to satisfy any of the metatheory we had in mind. At this point, our aim changed to the current aim of the paper: to pin down the problems in Verilog's simulation semantics that remain despite multiple previous projects having formalised the semantics.

Looking ahead, now having identified enough problems in the Verilog standard's description of Verilog's simulation semantics to turn previous work on Verilog formalisation into a formalisation compatible with Verilog practice, we hope that our future attempt at developing mechanised metatheory for the simulation semantics will end more positively than our first attempt. Ultimately, of course, the problems identified in this paper need to be addressed in the Verilog standard, such that the standard does not require patching to be usable.

\section{Data-Availability Statement}

The artefact~\cite{artefact} of this paper contains: (1) the small-scale systematic test suite we have developed and used to test how closed-source simulators handle the PREEMPT problem; (2) the source code of our new tool VV; (3) our modified version of Chen et al.'s artefact.

\bibliographystyle{ACM-Reference-Format}
\bibliography{paper}

\ifarxiv
\appendix
\section{Interleavings: Realistic Examples}%
\label{app:interleavings}


\paragraph{First realistic example} Consider the following variant of the \texttt{interleave3} module from Fig.~\ref{fig:interleave} where the module has been split into a synthesisable hardware model and a test bench:
\begin{figure}[h!]
\begin{minipage}[t]{0.41\textwidth}
\begin{minted}{verilog}
module interleaving(
 input logic a,
 output logic b,
 output logic c);

// First combinational logic block,
// assigns b twice
always_comb begin
 // First assignment of b
 b = 0;

 // Second assignment of b
 b = a;
end

// Second combinational logic block,
// depends on b
always_comb
 c = b;

endmodule
\end{minted}
\end{minipage}
\hfill
\begin{minipage}[t]{0.58\textwidth}
\begin{minted}{verilog}
module interleaving_tb;

logic a, b, c;

initial #1 a <= 1;

initial $monitor("a = %b, b = %b, c = %b", a, b, c);

interleaving interleaving(.a(a), .b(b), .c(c));

endmodule
\end{minted}
\end{minipage}
\end{figure}

\noindent
If arbitrary interleavings are allowed, the module \texttt{interleaving\_tb} can break in the same way as the module \texttt{interleave3}. Say we have \texttt{a} = \texttt{b} = \texttt{c} = \iver{'x} at the end of the first time slot. After the nonblocking-assignment event has been executed in the second time slot, the two \iver{always_comb} blocks can interleave in the same incorrect way as the corresponding blocks can in \texttt{interleave3}.



%

\paragraph{Second realistic example} Other realistic examples can be found in the test cases from the Icarus test suite that failed in Chen et al.'s evaluation (see in particular the file \texttt{scripts/data-race-cases.list} in their artefact~\cite{Chen23b}). One illustrative example is the test case \texttt{talu.v}, included below. With arbitrary interleavings, the test case can fail because the \iver{always} block in the \iver{alu} module depends on multiple data objects and the module can therefore, just like the modules \iver{interleave1} and \iver{interleave2} in Fig.~\ref{fig:interleave}, miss any number of writes to these data objects. 

%
\begin{minted}{verilog}
/* Copyright (C) 1999 Stephen G. Tell
 *
 * This program is free software; you can redistribute it and/or modify
 * it under the terms of the GNU General Public License as published by
 * the Free Software Foundation; either version 2, or (at your option)
 * any later version.
 *
 * This program is distributed in the hope that it will be useful,
 * but WITHOUT ANY WARRANTY; without even the implied warranty of
 * MERCHANTABILITY or FITNESS FOR A PARTICULAR PURPOSE.  See the
 * GNU General Public License for more details.
 *
 * You should have received a copy of the GNU General Public License
 * along with this software; see the file COPYING.  If not, write to
 * the Free Software Foundation, Inc., 59 Temple Place, Suite 330,
 * Boston, MA 02111-1307 USA
 */

/* talu - a verilog test,
 * illustrating problems I had in fragments of an ALU from an 8-bit micro
 */

module talu;
   reg error;

   reg [7:0] a;
   reg [7:0] b;
   reg cin;
   reg [1:0] op;

   wire cout;
   wire [7:0] aluout;

   alu alu_m(a, b, cin, op, aluout, cout);

   initial begin
       error = 0;

       // add
       op='b00; cin='b0; a='h0; b='h0;
       #2 if({cout, aluout} != 9'h000) begin
	   $display($time, " FAILED %b  %b %h %h  %b %h",
	            op, cin, a, b, cout, aluout);
	   error = 1;
       end

       // add1
       op='b01; cin='b0; a='h01; b='h01;
       #2 if({cout, aluout} != 9'h103) begin
	   $display($time, " FAILED %b  %b %h %h  %b %h",
	            op, cin, a, b, cout, aluout);
	   error = 1;
       end

       // and
       op='b10; cin='b0; a='h16; b='h0F;
       #2 if({cout, aluout} != 9'h006) begin
	   $display($time, " FAILED %b  %b %h %h  %b %h",
	            op, cin, a, b, cout, aluout);
	   error = 1;
       end
       op='b10; cin='b0; a='h28; b='hF7;
       #2 if({cout, aluout} != 9'h020) begin
	   $display($time, " FAILED %b  %b %h %h  %b %h",
	            op, cin, a, b, cout, aluout);
	   error = 1;
       end

       // genbit
       op='b11; cin='b0; a='h00; b='h03;
       #2 if({cout, aluout} != 9'h008) begin
	   $display($time, " FAILED %b  %b %h %h  %b %h",
	            op, cin, a, b, cout, aluout);
	   error = 1;
       end
       op='b11; cin='b0; a='h00; b='h00;
       #2 if({cout, aluout} != 9'h001) begin
	   $display($time, " FAILED %b  %b %h %h  %b %h",
	            op, cin, a, b, cout, aluout);
	   error = 1;
       end

       /* tests are incomplete - doesn't compile yet on ivl */

       if(error == 0)
	   $display("PASSED");

       $finish;
   end

endmodule

/* fragments of an ALU from an 8-bit micro
 */

module alu(Aval, Bval, cin, op, ALUout, cout);
   input [7:0]  Aval;
   input [7:0]  Bval;
   input        cin;
   input [1:0]  op;
   output       cout;
   output [7:0] ALUout;

   reg          cout;
   reg [7:0]    ALUout;

   always @(Aval or Bval or cin or op) begin
      case(op)
	2'b00 : {cout, ALUout} = Aval + Bval;
	2'b10 : {cout, ALUout} = {1'b0, Aval & Bval};

// C++ compilation troubles with both of these:
	2'b01 : {cout, ALUout} = 9'h100 ^ (Aval + Bval + 9'h001);
	2'b11 : {cout, ALUout} = {1'b0, 8'b1 << Bval};

//	2'b01 : {cout, ALUout} = 9'h000;
//	2'b11 : {cout, ALUout} = 9'h000;
      endcase
   end // always @ (Aval or Bval or cin or op)

endmodule
\end{minted}

\section{Interleavings: Previous Work}%
\label{app:executions}


Since we will be referring to the Verilog semantics of Chen et al.~\cite{Chen23} and Meredith et al.~\cite{Meredith10} repeatedly in this appendix, we refer to the two semantics as $\lambda$-Verilog and K-Verilog, respectively, in this appendix. Both semantics have state-explorer tools, allowing us to explore all reachable behaviour of the semantics. $\lambda$-Verilog can also be run as an interpreter, where we can supply a ``seed'' to control which schedule is used during execution. For modules that we run in $\lambda$-Verilog and K-Verilog, we must use \iver{always @(*)} instead of \iver{always_comb} because $\lambda$-Verilog and K-Verilog are formalisations of Verilog-2005, not SystemVerilog. Here, the only difference between \iver{always @(*)} and \iver{always_comb} is that the latter automatically runs once in the first time slot.

\paragraph{PREEMPT} We now show that the module \iver{interleave3} from Fig.~\ref{fig:interleave} in the main text has problematic interleavings in both $\lambda$-Verilog and K-Verilog. We use the following variant of the module:
\begin{minted}{verilog}
module interleave3_observable;

// We use an array here since
// K-Verilog does not support
// X values
reg[1:0] a, b;

always @(*)
 b = a;

initial begin
 a = 1;
 a = 2;
end

initial $monitor("a = %b, b = %b", a, b);

endmodule
\end{minted}

\noindent
Running the $\lambda$-Verilog interpreter with seed 38 gives the following output:
\begin{verbatim}
> lv -ci interleave3_observable.v --seed=38 -o tmp.lv
a = 10, b = 01
\end{verbatim}
The state-space explorer of K-Verilog confirms that the same outcome is possible in the K-Verilog semantics by reporting the following two reachable outputs:
\begin{verbatim}
a = 10, b = 01
\end{verbatim}
and
\begin{verbatim}
a = 10, b = 10
\end{verbatim}

\paragraph{NB\_MIX} To show that both $\lambda$-Verilog and K-Verilog allow for NBA events to mix with other events, consider the following version of the module \texttt{nbinterleave2} from~Fig.~\ref{fig:nbproblem}:
\begin{minted}{verilog}
module nbinterleave2;

reg[1:0] a;

initial begin
 a <= 1;
 a <= 2;
end

always @(*)
 $display("a = %b", a);

endmodule
\end{minted}

\noindent
$\lambda$-Verilog has interleavings of the NBA events and the display process:
\begin{verbatim}
> lv -ci nbinterleave2.v --seed=12 -o tmp.lv
a = 01
a = 10
\end{verbatim}
Similarly, the K-Verilog state-space explorer reports the following as reachable outputs:
\begin{verbatim}
a = 1
\end{verbatim}
and
\begin{verbatim}
a = 10
\end{verbatim}
and
\begin{verbatim}
a = 1
a = 10
\end{verbatim}

\paragraph{NB\_ORDER} Since both $\lambda$-Verilog and K-Verilog mix NBA events with other events, we can easily observe that none of them reorder NBA events. To demonstrate this, consider the following version of the module \texttt{nbinterleave3} from~Fig.~\ref{fig:nbproblem}:
\begin{minted}{verilog}
module nbinterleave3;

reg a, b;

initial begin
 a <= 1;
 b <= 1;
end

always @(*)
 $display("a = %b, b = %b", a, b);

endmodule
\end{minted}

\noindent
The state-space explorer of $\lambda$-Verilog reports:
\begin{verbatim}
> lv -cx nbinterleave3.v -o tmp.lv | grep -v '^Heap'
a = 1, b = x
a = 1, b = x
a = 1, b = 1
a = 1, b = 1
a = 1, b = 1
a = 1, b = 1
\end{verbatim}
For K-Verilog, the state-space explorer reports the following three reachable output states:
\begin{verbatim}
a = 1, b = 0
\end{verbatim}
and
\begin{verbatim}
a = 1, b = 1
\end{verbatim}
and
\begin{verbatim}
a = 1, b = 0
a = 1, b = 1
\end{verbatim}
That is, for both semantics, \texttt{b} is never assigned before \texttt{a}.

\section{VV}%
\label{app:vv}

\begin{figure}[b]
\begin{minipage}[t]{0.54\textwidth}
\begin{Verbatim}[fontsize=\small]
type value
 = BitTrue
 | BitFalse
 | BitX
 | BitZ

type rec event
 = EventContUpdate(int, value)
 | EventBlockUpdate(int, string, value)
 | EventNBA(string, value)
 | EventEvaluation(int)
 | EventDelayedEvaluation(int)
 | Events(array<event>)

type time_slot = {
 active: array<event>,
 inactive: array<event>,
 nba: array<event> }
\end{Verbatim}
\end{minipage}
\begin{minipage}[t]{0.45\textwidth}
\begin{Verbatim}[fontsize=\small]
type proc_running_state
 = ProcStateFinished
 | ProcStateRunning
 | ProcStateWaiting

type proc_state = {
 pc: int,
 state: proc_running_state }

type state = {
 // [...]
 env: Belt.Map.String.t<value>,
 proc_env: array<proc_state>,
 cont_env: array<value>,
 queue: array<(int, time_slot)>,
 monitor: option<(string, /* ... */)> }
\end{Verbatim}
\end{minipage}
\caption{The simulation-state representation \texttt{state} as implemented in VV. The code shown here is ReScript code~\cite{rescript}, an OCaml dialect of JavaScript, which is the language we have used to implement VV. We have made some small simplifications to the code shown here to make it easier to read. We mention two of the simplifications. First, the full \texttt{state} data type contains a few more fields not interesting enough to mention here and are therefore omitted in the presentation here (\texttt{// [...]}). Second, the actual \texttt{event} data type contains \texttt{event\_id}s as well, but they are only used for the GUI of the tool and do not~affect~any~behaviour.}%
\label{fig:state}                                
\end{figure}

In this appendix, we discuss additional implementation details of VV. We implemented VV from scratch because there is little overlap between existing simulators and VV. Existing simulators are batch tools designed for debugging real-world hardware designs, whereas VV is an interactive tool designed for visualising the Verilog's simulation semantics. Simulation speed is the main driving force behind the design of existing simulators, whereas for VV performance is largely unimportant. Even at the expense of performance, to be an accurate visualisation of the standard's description of Verilog's simulation semantics, VV must be in an as simple and direct correspondence with the Verilog standard as possible. E.g., the event queue maintained by VV must be exactly as described by the Verilog standard rather than implemented for performance. Moreover, in VV, the full behaviour of Verilog must be exposed, e.g., all event schedules must be exposed. In this respect, VV has more in common with explicit-state model checking than traditional simulators, except that VV is driven by human rather than machine.



\paragraph{Simulation-state representation} We now discuss the simulation-state representation of VV, which is the most interesting part of the implementation of VV. The standard defines the high-level structure of the Verilog event queue but does not make explicit how simulators and other Verilog tools should represent the rest of the simulation state. Fig.~\ref{fig:state} gives the top-level state structure of VV, called \texttt{state}, which contains: the state of all variables and nets (\texttt{env}), the state of all procedural processes (\texttt{proc\_env}), the state of all continuous assignments/drivers (\texttt{cont\_env}), the state of the event queue (\texttt{queue}), and, optionally, a monitor (\texttt{monitor}).

\paragraph{Process representation} Procedural processes, stored in the field \texttt{proc\_env}, are similar to software processes: the data type \texttt{proc\_state} therefore contains the current PC (``program counter'') and the current running state of the process (\texttt{proc\_running\_state}). The field \texttt{cont\_env} stores all continuous assignments: each continuous assignment induces a driver process that only needs to keep track of current value of the expression of the assignment.

\paragraph{Event-queue representation} Both the data type of the event queue field \texttt{queue} and the data type \texttt{time\_slot} can be directly read off the reference algorithm. The event queue field \texttt{queue} is represented by a series of time slots of type \texttt{time\_slot} indexed by the simulation time (an \texttt{int}) of the time slot. Each time slot (i.e., the type \texttt{time\_slot}) consists of the regions \texttt{active}, \texttt{inactive}, and \texttt{nba}. In our target subset of Verilog, no field for the observed region is needed in the time slot data type since only monitor invocations are scheduled in the observed region and monitors are only ever scheduled for all future time slots, not for a specific time slot: monitors are instead represented using the \texttt{monitor} field in the \texttt{state} data type.

\paragraph{Event representation} The reference algorithm does not dictate the exact structure of events, instead, it only mentions that there are two types of events (see the \texttt{execute\_region} function): ``update'' events and ``evaluation'' events. In VV, the two categories are refined into the \texttt{event} data type containing six event types. The numbers of event types and the structure of each event type are not canonical, instead, the two are largely a consequence of how other components of the interpreter are implemented; i.e., other interpreters/semantics might sensibly end up with other refinements. The event types \texttt{EventContUpdate} and \texttt{EventBlockUpdate}, respectively, represent a continuous assignment update and an update scheduled by a procedural blocking assignment, where the \texttt{int} in the event types is the index of the continuous assignment/procedural process. The two event types \texttt{EventEvaluation} and \texttt{EventDelayedEvaluation} both represent the start of execution of procedural processes, and are separated only because of a small (and not particularly interesting) edge case. The event types \texttt{EventNBA} and \texttt{Events} are used to represent nonblocking assignments: executing a nonblocking assignment schedules an \texttt{EventNBA} event in the relevant NBA region, and when the NBA events are later moved to the active region, following Meredith et al.'s trick to enforce \orderall (discussed in Sec.~\ref{sec:intramodule-problems}), the order between them is preserved by grouping them inside an \texttt{Events} event. The \texttt{nba} field of \texttt{time\_slot} will only ever contain \texttt{EventNBA} events and, similarly, \texttt{Events} events will only ever contain \texttt{EventNBA} events.




\else
\fi


\end{document}